\documentstyle[graphicx,natbib,epsfig,amsmath,amssymb,traditabstract]{aa} 

\newcommand{\msun}{\,\hbox{$M_{\odot}$}}

\newcommand{\kms}{\,\hbox{\hbox{km}\,\hbox{s}$^{-1}$}}

\newcommand{\reff}{\mbox{$R_{\rm eff}$}}

\newcommand{\htwo}{\hbox{$\rm{H_ 2}$}}
\newcommand{\ha}{\,\hbox{${\rm H\alpha}$}}

\newcommand{\hi}{\,\hbox{\ion{H}{I}}}
\newcommand{\nii}{\,\hbox{[\ion{N}{II}]}}

\newcommand{\um}{\,\hbox{$\mu$m}}

\newcommand{\cooz}{\hbox{$^{12}\rm CO(1-0)$}}
\newcommand{\coto}{\hbox{$^{12}\rm CO(2-1)$}}
\newcommand{\her}{\hbox{\it Herschel}}

\begin{document}
 
 \setlength{\parskip}{0.5mm plus 0.0mm minus0.0mm}

   \title{Survival of molecular gas in Virgo's hot intracluster medium: \\ CO near M86}

   \subtitle{}

   \author{K. M. Dasyra\inst{1} 
          \and
	 F.  Combes\inst{1} 
	  \and
	  P. Salom\'e\inst{1}
	  \and
	  J. Braine\inst{2,3}
	  }

   \institute{
             Observatoire de Paris, LERMA, CNRS, 61 Avenue de l'Observatoire, F-75014, Paris, France 
             \and
             Univ. Bordeaux, Laboratoire d'Astrophysique de Bordeaux, UMR 5804, F-33270, Floirac, France
             \and
             CNRS, LAB, UMR 5804, F-33270, Floirac, France
             }

   \date{}

    \abstract  
    { We carried out \cooz\ and \coto\ observations of 21 different regions in the vicinity of M86, NGC4438, and along the 120 kpc-long,  \ha 
    -emitting filamentary trail that connects them, aiming to test whether molecular gas can survive to be transferred from a spiral to an elliptical 
    galaxy in Virgo's 10$^7$K intracluster medium (ICM). We targeted \ha -emitting regions that could be associated with the interface between 
    cold molecular clouds and the hot ionized ICM. The data, obtained with the 30m telescope of the Institut de Radioastronomie Millim\'etrique, led to 
    the detection of molecular gas close to M86. CO gas with a recession velocity that is similar to that of the stars, $-$265\kms , and with a corresponding 
    \htwo\ mass of  2$\times$10$^7$\msun , was detected $\sim$10 kpc southeast of the nucleus of M86, near the peak of its \hi\  emission. We argue 
    that it is possible for this molecular gas either to have formed in situ from \hi , or to have been stripped from NGC4438 directly in molecular form. In 
    situ formation is nonetheless negligible for the 7$\times$10$^6$\msun\ of gas detected at 12:26:15.9$+$12:58:49, at $\sim$10 kpc northeast of M86, 
    where no (strong) \hi\  emission is present. This detection provides evidence for the survival of molecular gas in filaments for timescales of $\sim$100 
    Myr. An amount equivalent to 5$\times$10$^7$\msun\ of \htwo\ gas that could be lost to the ICM or to neighboring galaxies was also discovered in the 
    tidal tail northwest of NGC4438. A scenario in which gas was alternatively brought to M86 from NGC4388 was also examined but it was considered 
    unlikely because of the non-detection of CO below or at the H\,I stream velocities, 2000-2700\kms .}

   \titlerunning{CO in Virgo's hot intracluster medium}
   \authorrunning{Dasyra et al.}
   
   \maketitle


\section{Introduction}
\label{sec:intro}

Stripping of gas during galaxy collisions can replenish the gas reservoirs of elliptical galaxies and reinitiate star formation 
in them, in particular in overdense cluster environments. Cooling flows can further add to the deposition of external gas in cluster
ellipticals \citep[see][and references therein]{fabian94}. The growth of these ellipticals is regulated by the interplay of gas cooling and 
heating along gas filaments that are associated with either mergers or cooling flows and that have often been discovered in the \ha\ 
and \nii\ images of local clusters \citep{cowie83,heckman89,conselice01,kenney08}. This interplay manifests itself and can be examined 
through the detection of multi-phase gas emission lines in the same clouds. Indeed, warm atomic gas, warm molecular gas, and cold 
molecular gas clouds have been found to coexist in spatially resolved filaments around NGC1275 (Perseus A), where they were 
detected through \ha , \htwo , and CO emission \citep{conselice01,salome06,salome11,johnstone07}. 

Several processes, however, occur in cluster environments that can impede the transfer of gas from one galaxy to another. Ram pressure 
stripping \citep{gunn72}, for example, can disperse the diffuse atomic gas into the intracluster medium (ICM). Even though the ram 
pressure drag is negligible for the motions of the dense molecular gas clouds \citep{nulsen82,kenney89}, it can eventually destroy them by 
depleting the atomic gas reservoir from which they reform \citep{crowl05}. The molecular clouds are also several orders of magnitude cooler 
than their environment and can be easily destroyed by X-rays if they are not self-shielded. The ICM X-ray spectrum often peaks in the 
range 0.7-2.0\,keV, indicating temperatures of $\sim$$10^7$K \citep[e.g.,][]{rangarajan95,machacek04,fabian06,randall08,tamura09}.


\begin{figure*}
\begin{center}
\includegraphics[width=18.8cm]{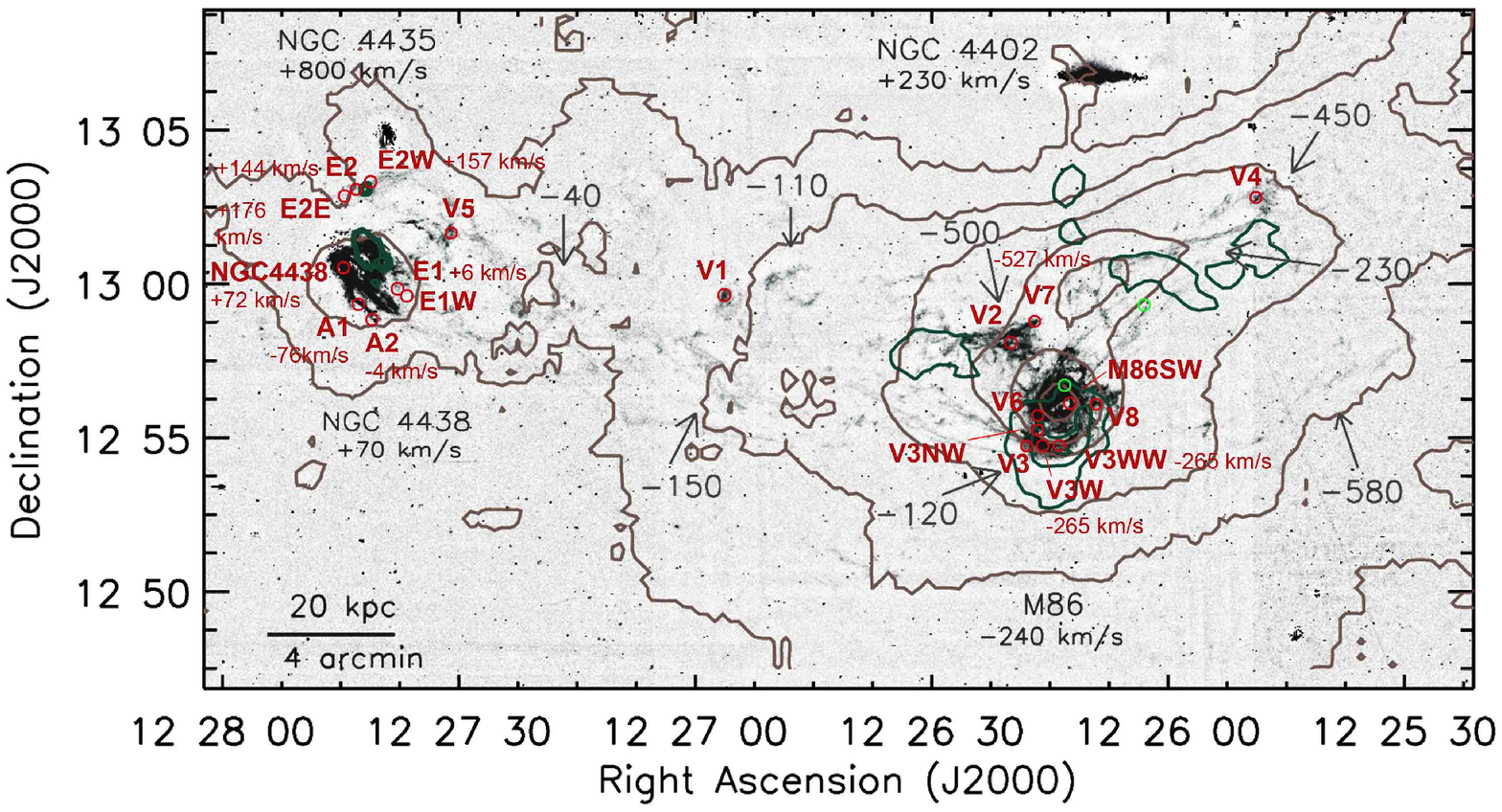} 
\caption{ Finding chart of our observations. The positions of previous CO observations (\citealt{wiklind95}; \citealt{braine97})
are marked with light green circles. HI line contours (dark green) and X-ray contours (brown) are plotted over the narrow
band \ha $+$\nii\ image of Virgo's core (adapted from \citealt{kenney08}). The velocities appearing in gray correspond to those
of the \ha\ gas as measured by the same authors. In red, we mark the \ha\ clouds that we observed searching for associated CO 
emission. Upon detection, we also indicate the velocity of the molecular gas traced by the \cooz\ line. In addition to these 20 regions, 
we also observed at 12:27:30.0+13:12:12.0, northwest of NGC4435, in the clouds that \citet{cortese10a} denote as plume.
}
\label{fig:finding_chart}
\end{center}
\end{figure*}

Evidence that molecular clouds do nonetheless survive or reform in $\gtrsim$10$^7$K intracluster and intragalactic media 
has been found. \citet{braine00} discovered CO in a tidal tail dwarf galaxy of Arp105, in the X-ray emitting medium of Abell 1185 
\citep{mahdavi96}. Because the CO-emitting regions spatially coincided and had comparable kinematic properties with the 
H\,I-emitting regions, \citet{braine00} proposed in situ formation of molecules. In addition to the molecular gas detected in the 
cooling-flow-related filaments around NGC1275 in Perseus \citep{salome06,salome11,johnstone07},  
\citet{vollmer08} found extraplanar CO near NGC4522 in Virgo,  in regions where atomic gas was brought by ram pressure. In 
ESO137-001, located in Abell 3627, a $\sim$$10^7$K ram-pressure-driven gas tail has active star formation and \htwo\ 
emission  \citep{sun10, sivanandam10}.  In the compact group of Stephan's Quintet, \htwo\ molecules not  only survive in a medium 
of comparable temperature \citep{trinchieri05,osullivan09}, but they are mainly responsible for the gas cooling instead of the 
\hbox{X-rays \citep{cluver10}}.

An excellent system to test whether molecular gas could be directly brought to an elliptical galaxy, under extreme conditions, 
from spiral(s) interacting with it is in the Virgo cluster. The dust clumps that are detected in the giant elliptical galaxy M86 
(also known as NGC4406) could have originated from its past interaction with its neighboring spiral galaxy NGC4438
\citep{cortese10b,gomez10}. Dust and gas could have been captured  through ram pressure stripping of the interstellar medium 
(ISM) of NGC4438 by that of M86 \citep{kenney08}, through dynamical instabilities, or through both processes. The H\,I content 
deficit of NGC4438 relative to its own CO emission confirms that ram pressure has acted on its ISM \citep{vollmer05,vollmer09,
hota07}. Dynamical perturbations continue to distort NGC4438, as indicated by the presence of stars and gas in its northern 
tidal tail toward NGC4435 \citep{combes88}. The interaction between NGC4438 and M86 is confirmed by a spectacular, 
$\sim$120\,kpc-long, \ha -emitting tidal bridge that connects them \citep{kenney08}. The smooth \ha\ velocity gradient along 
this bridge convincingly places it to the ICM of Virgo, which is rich in X-ray emission \citep{forman79,rangarajan95,machacek04,
liu05,randall08}. Since the X-ray heating and the ram pressure disperse the ICM atomic gas, a significant fraction of the \ha\ emission 
could originate from the ionized gas in the outer layers of molecular clouds \citep{ferland09}. This makes the \ha\ clumps in the 
NGC4438-M86 bridge prime targets for testing whether star formation can be initiated from gas that has either been brought from 
NGC4438 to M86 in the molecular state, or from molecules that recombined in situ in M86 after being transferred in the atomic state.

We obtained mm data with the 30\,m telescope of the Institut de Radioastronomie Millim\'etrique (IRAM) to test for the presence of 
molecular gas in some of the \ha\ filaments that are embedded in Virgo's hot ICM. We focused on \ha\ emission peaks, aiming to 
determine whether the warm atomic gas emission could be associated with cold gas clouds, to compute the amount of molecular gas 
and potential star formation rate (SFR) in M86, and to test the gas origin by examining whether the molecules could have been 
transferred there or formed in situ. We assume a distance of 17.5 Mpc to Virgo throughout this work \citep{mei07}. 


\begin{figure*}
\scalebox{0.93}{
\rotatebox{90}{~~~~~~~~T$_{mb}$ (K)}
\includegraphics[width=5.8cm]{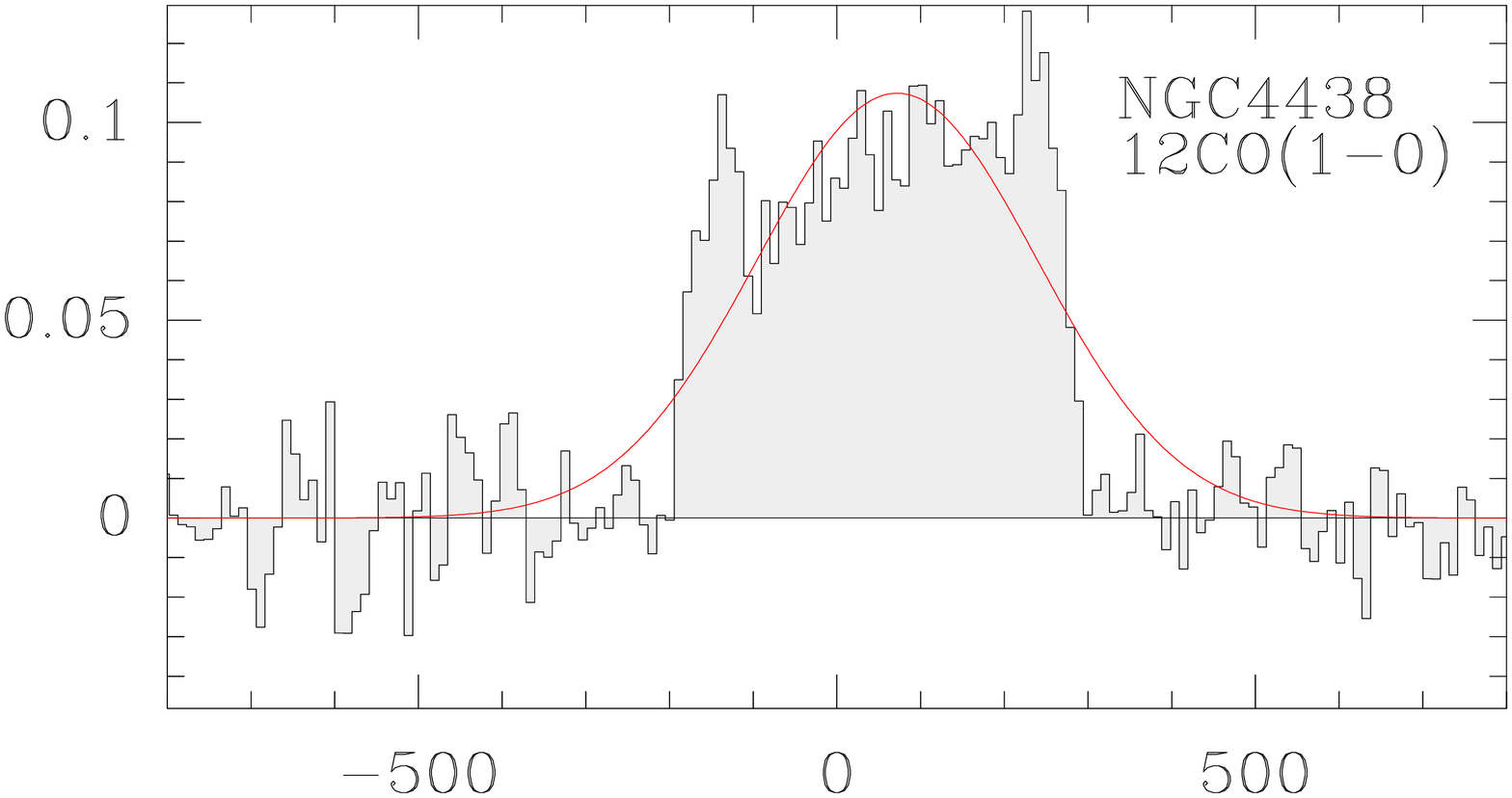}
\includegraphics[width=5.8cm]{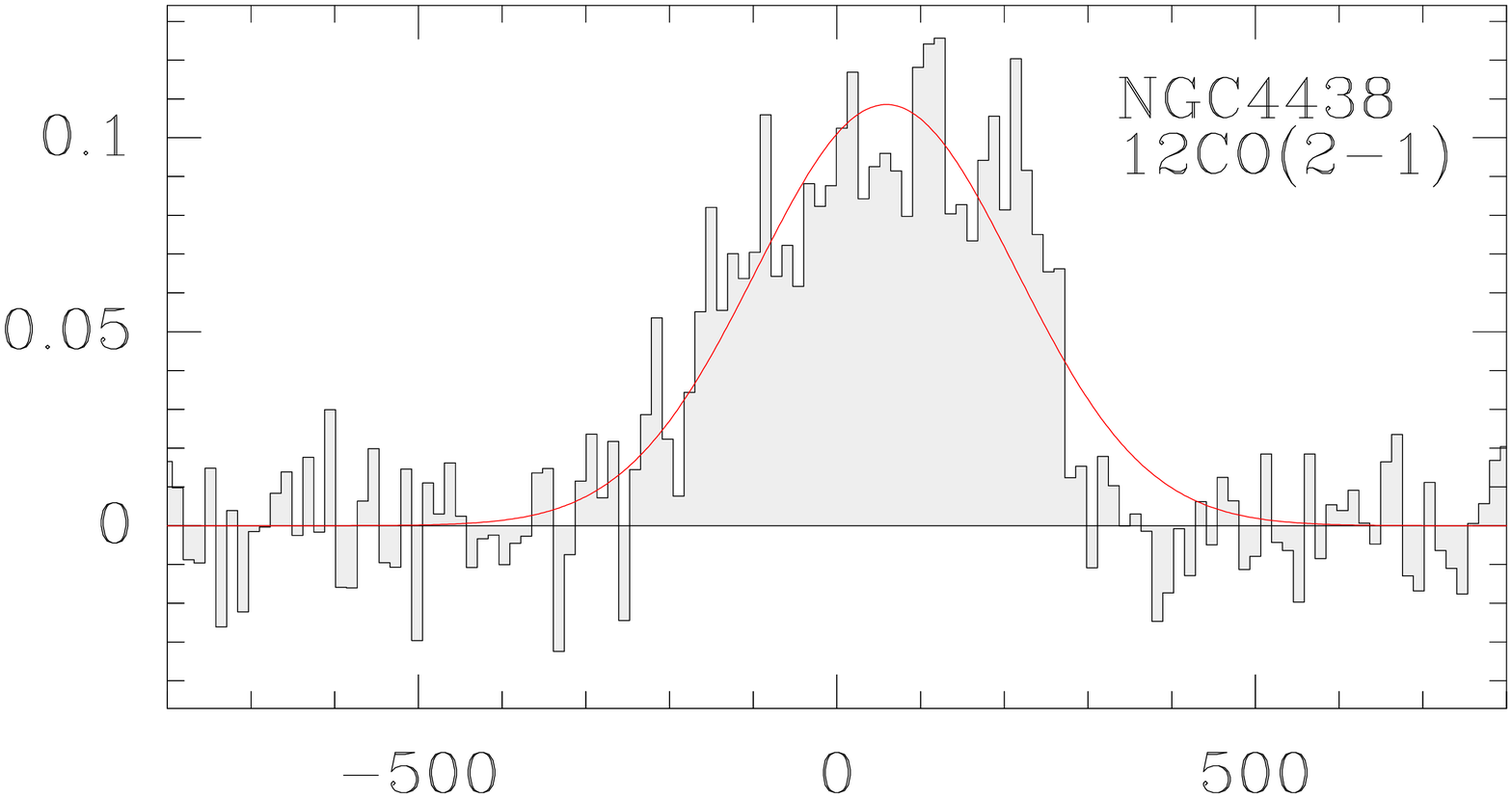}
\includegraphics[width=5.8cm]{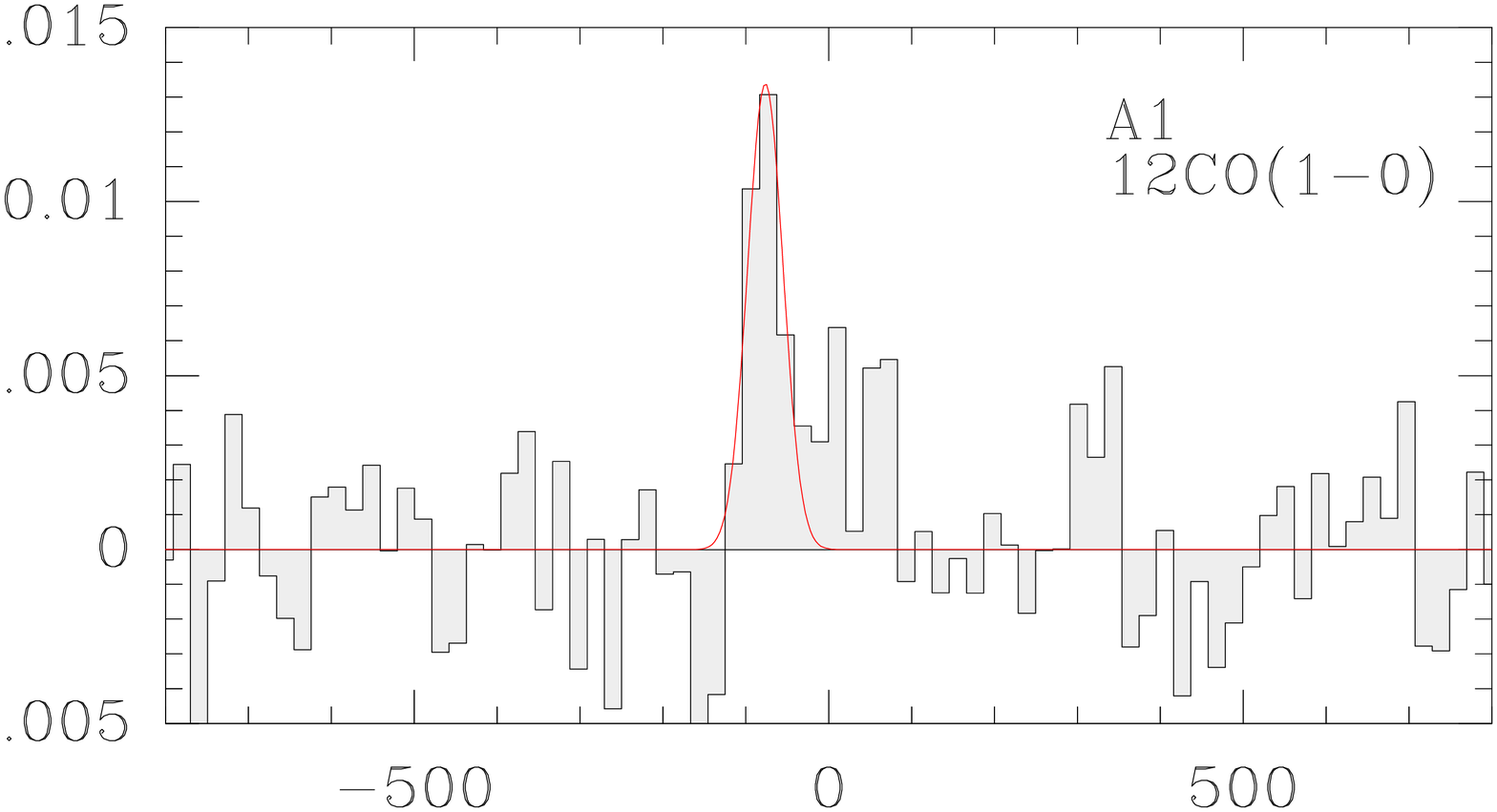} 
} \\
\scalebox{0.93}{
\rotatebox{90}{~~~~~~~~T$_{mb}$ (K)}
\includegraphics[width=5.8cm]{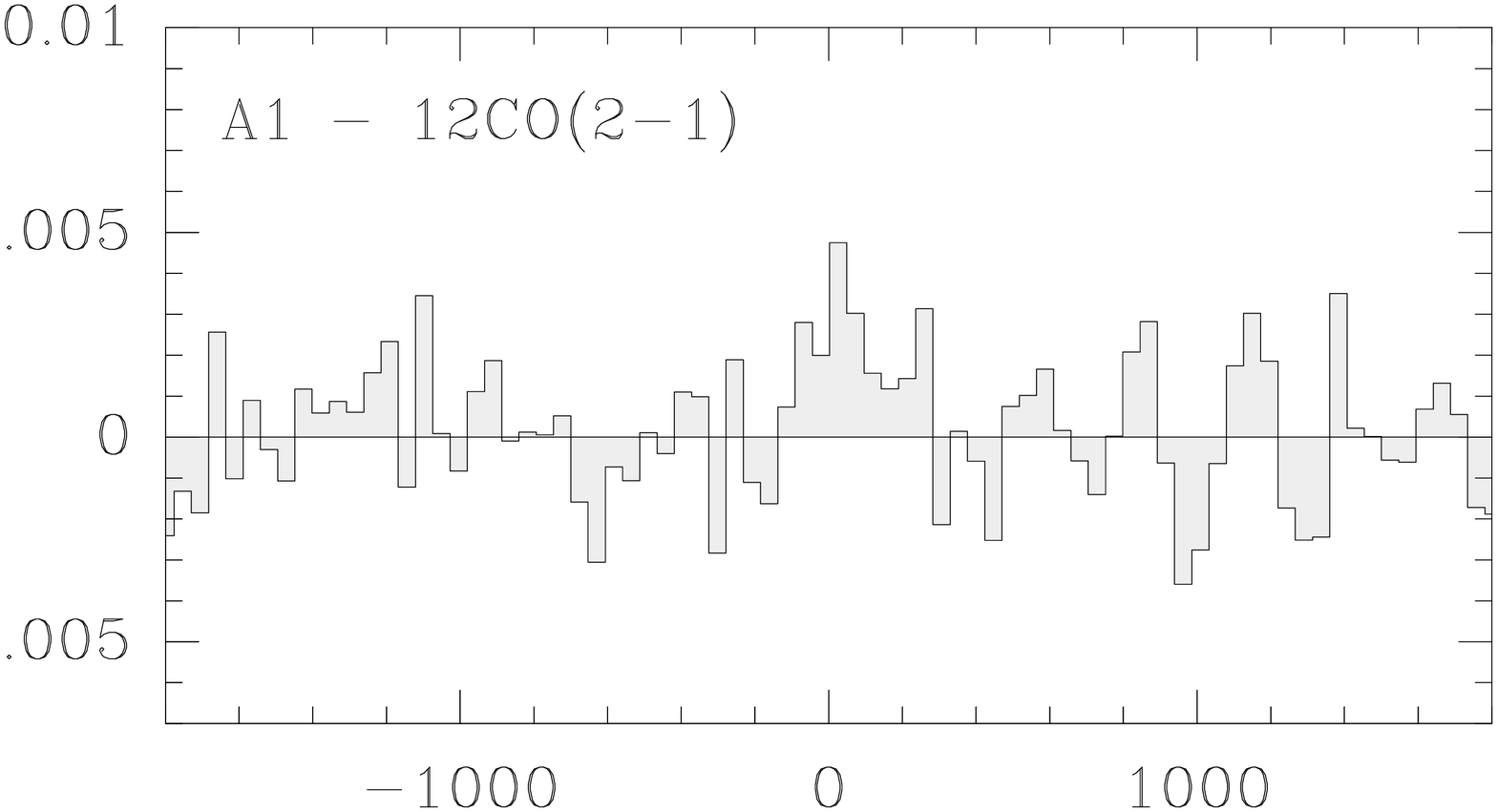}
\includegraphics[width=5.8cm]{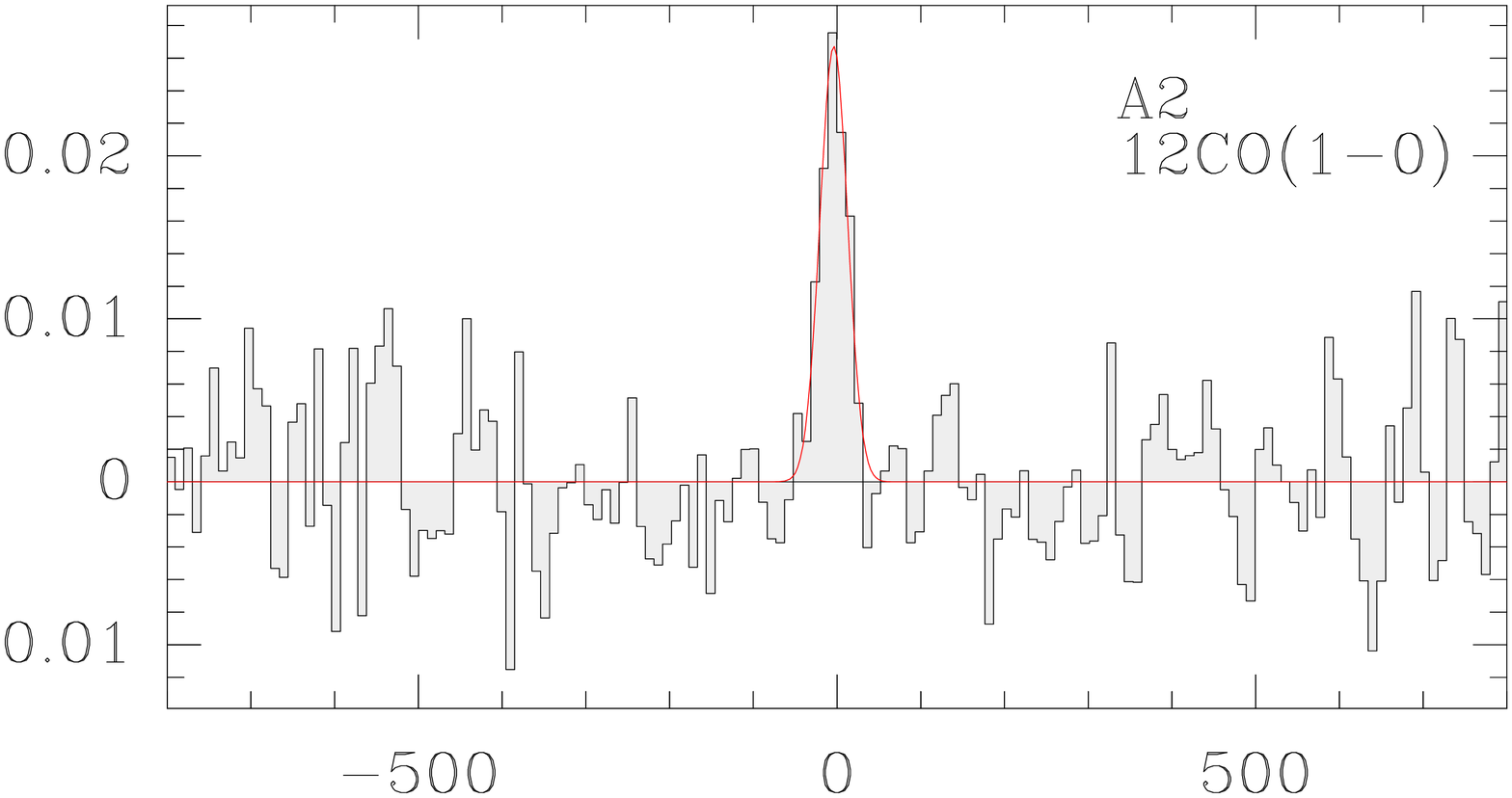}
\includegraphics[width=5.8cm]{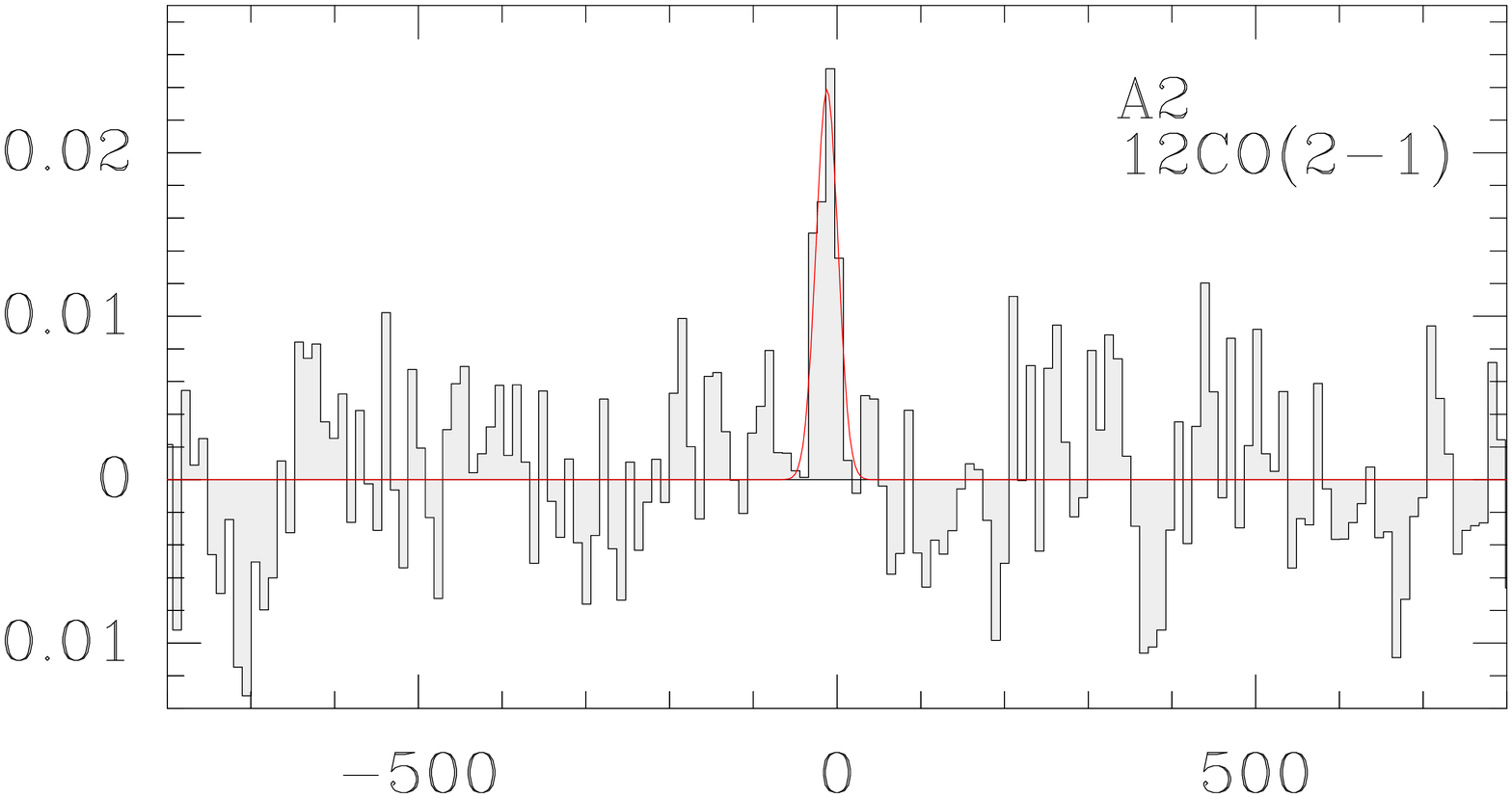}
} \\
\scalebox{0.93}{
\rotatebox{90}{~~~~~~~~T$_{mb}$ (K)}
\includegraphics[width=5.8cm]{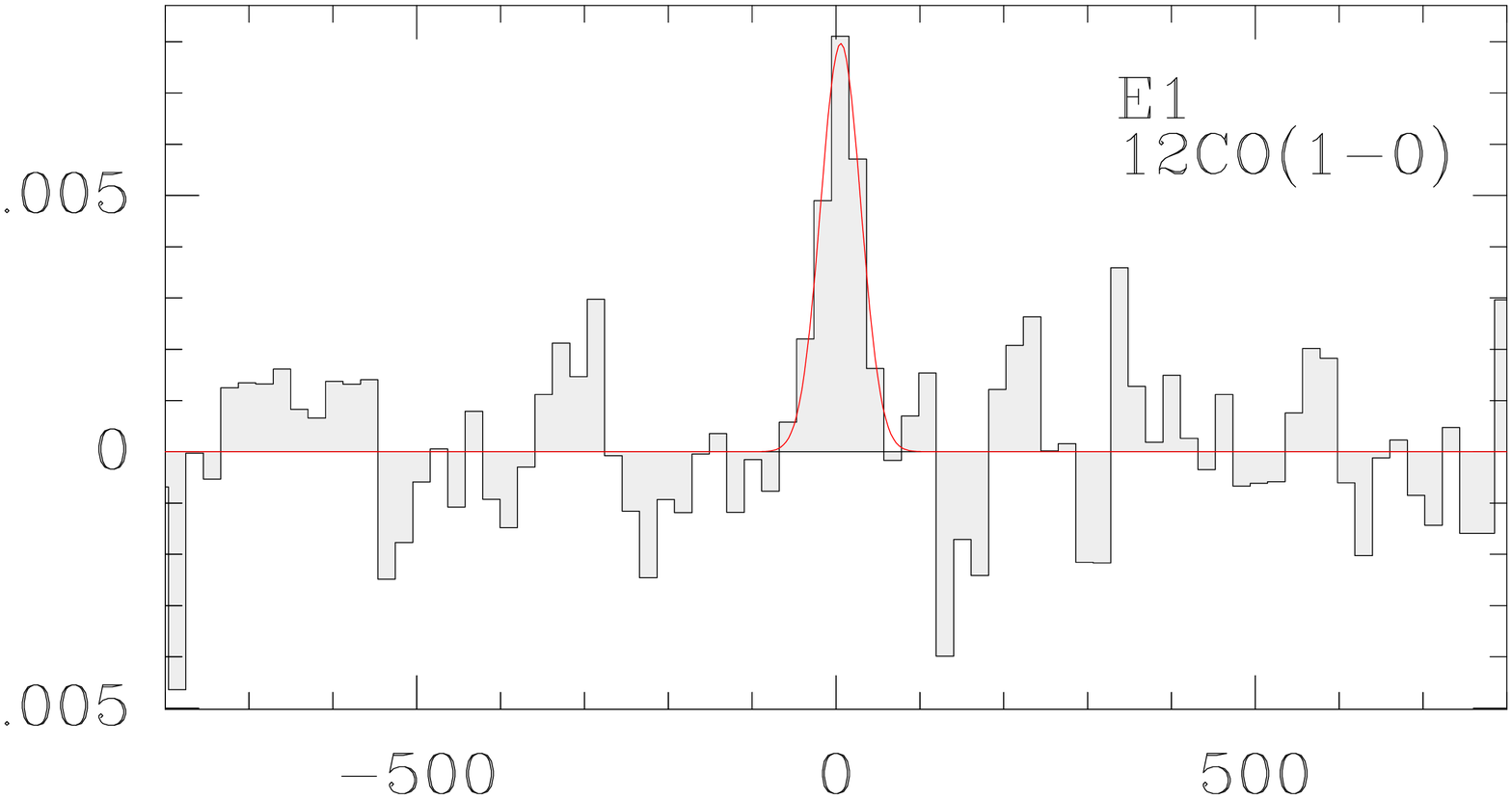}
\includegraphics[width=5.8cm]{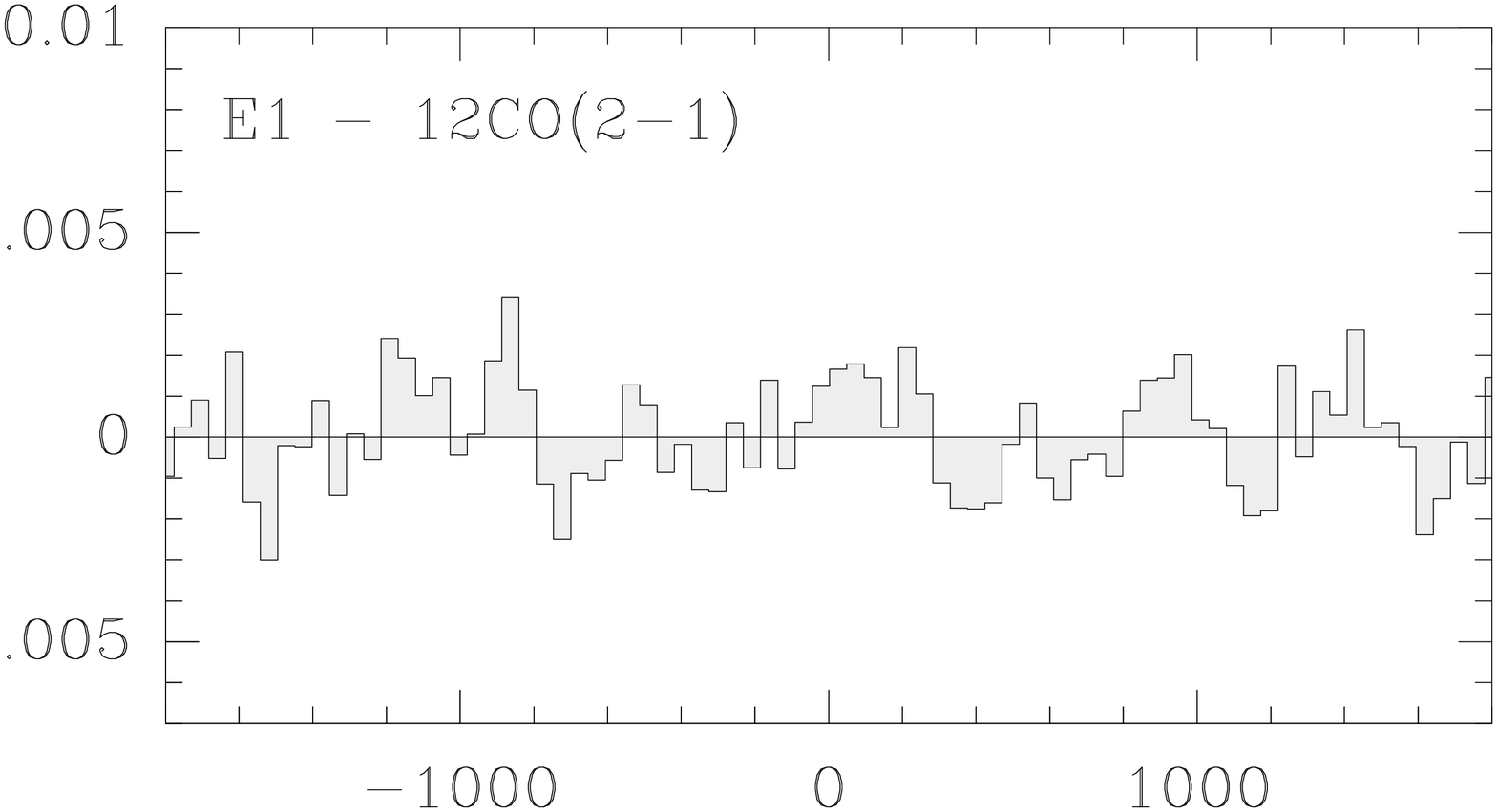}
\includegraphics[width=5.8cm]{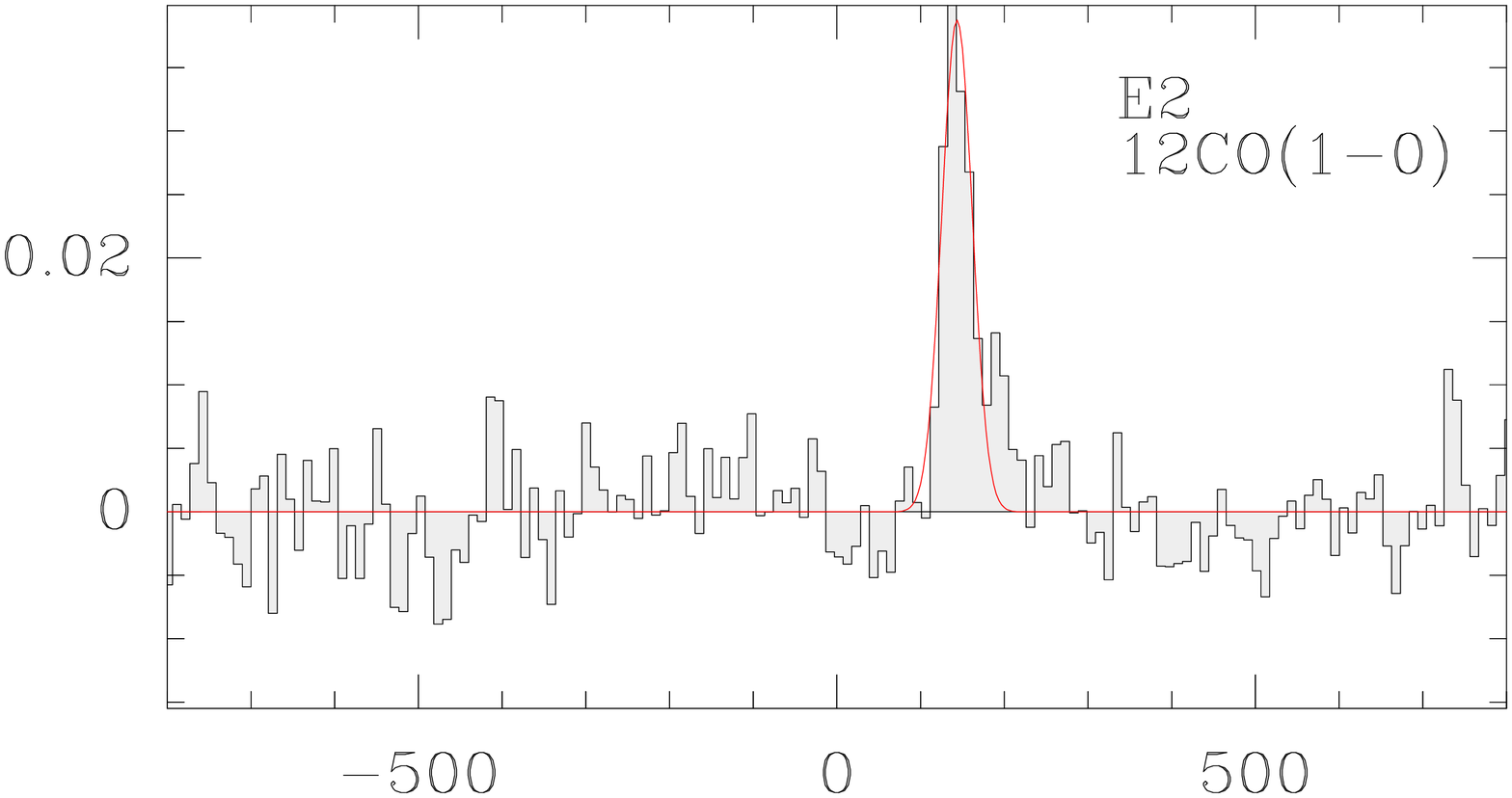}
} \\
\scalebox{0.93}{
\rotatebox{90}{~~~~~~~~T$_{mb}$ (K)}
\includegraphics[width=5.8cm]{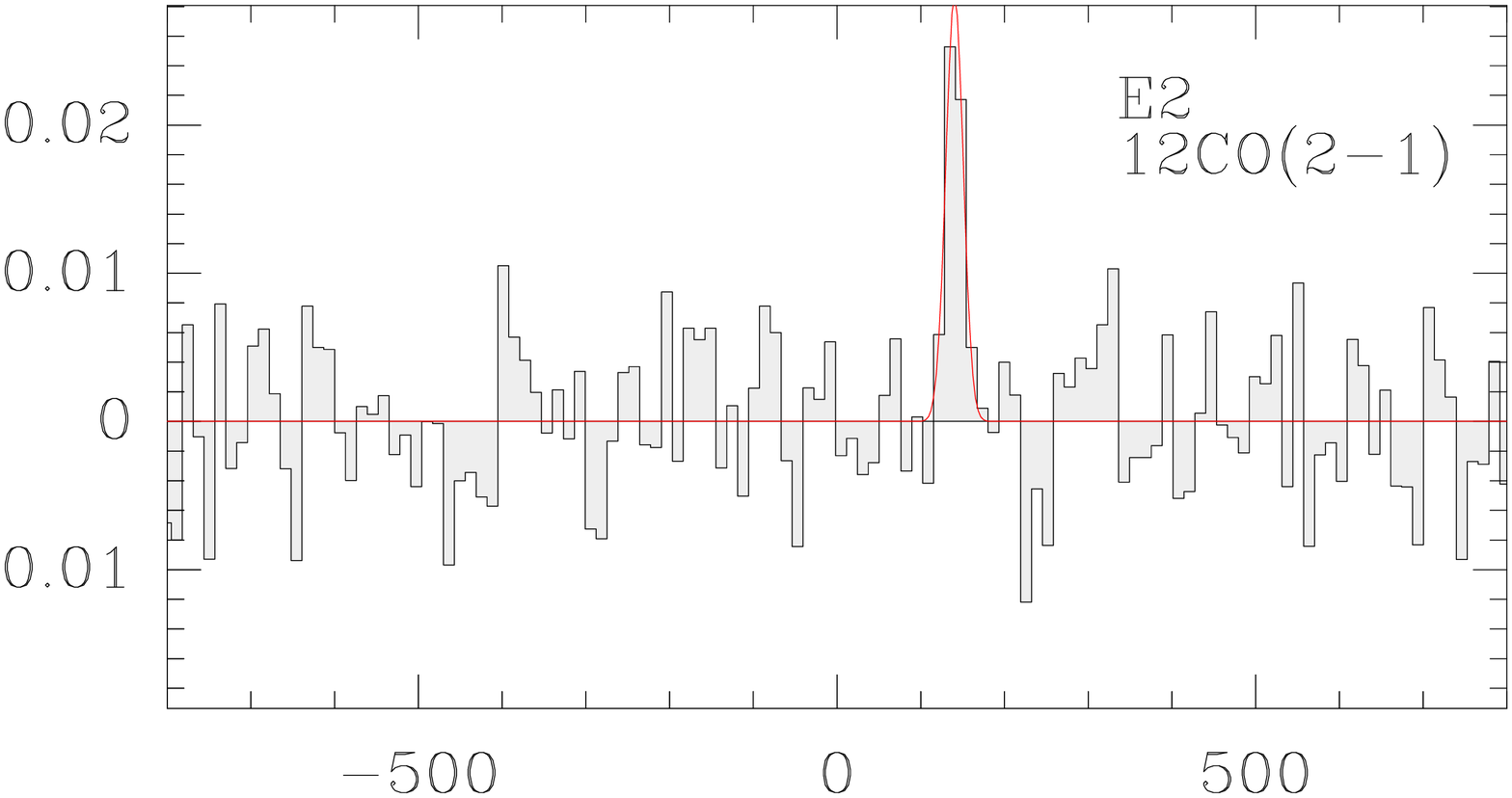}
\includegraphics[width=5.8cm]{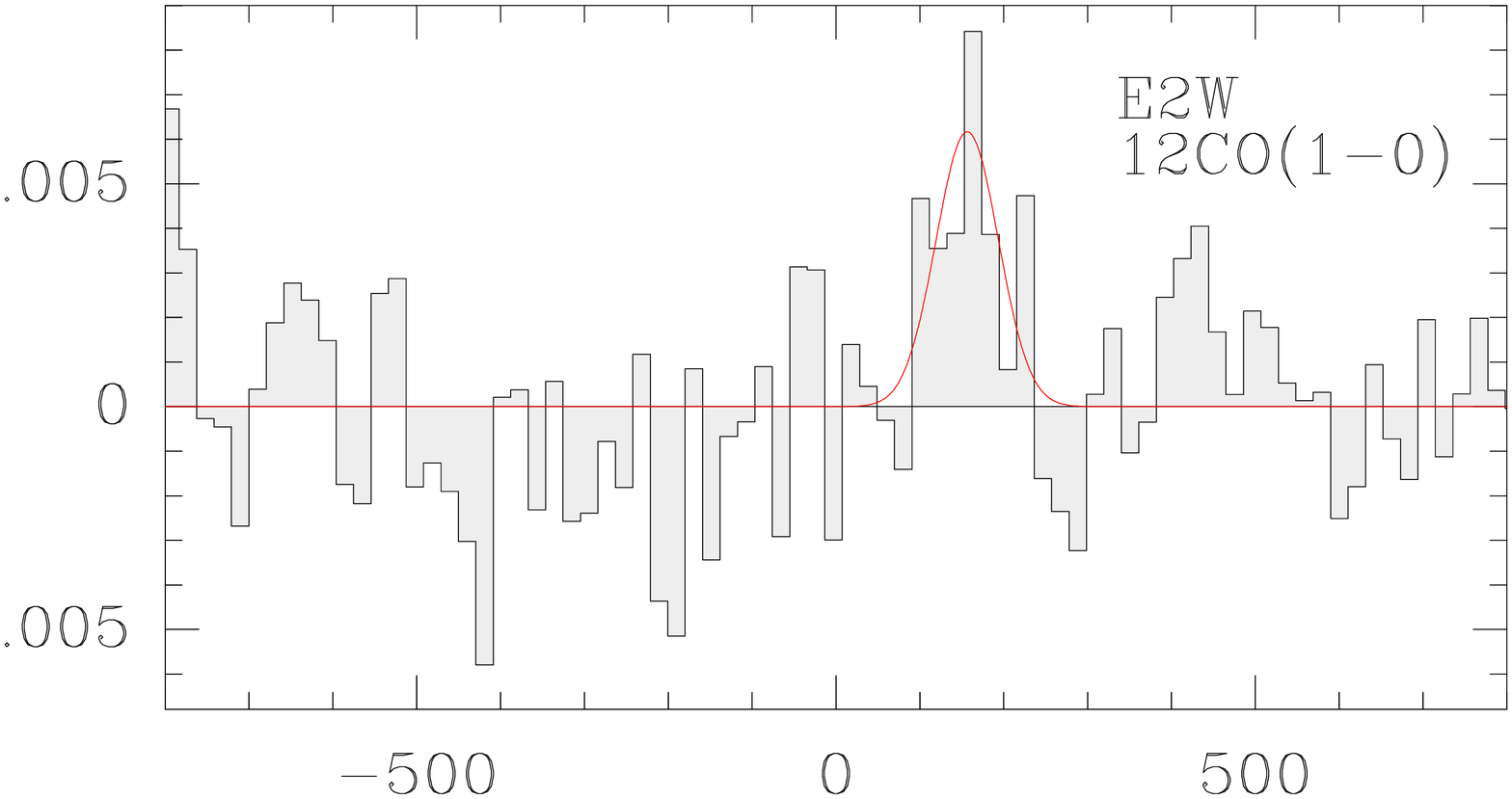}
\includegraphics[width=5.8cm]{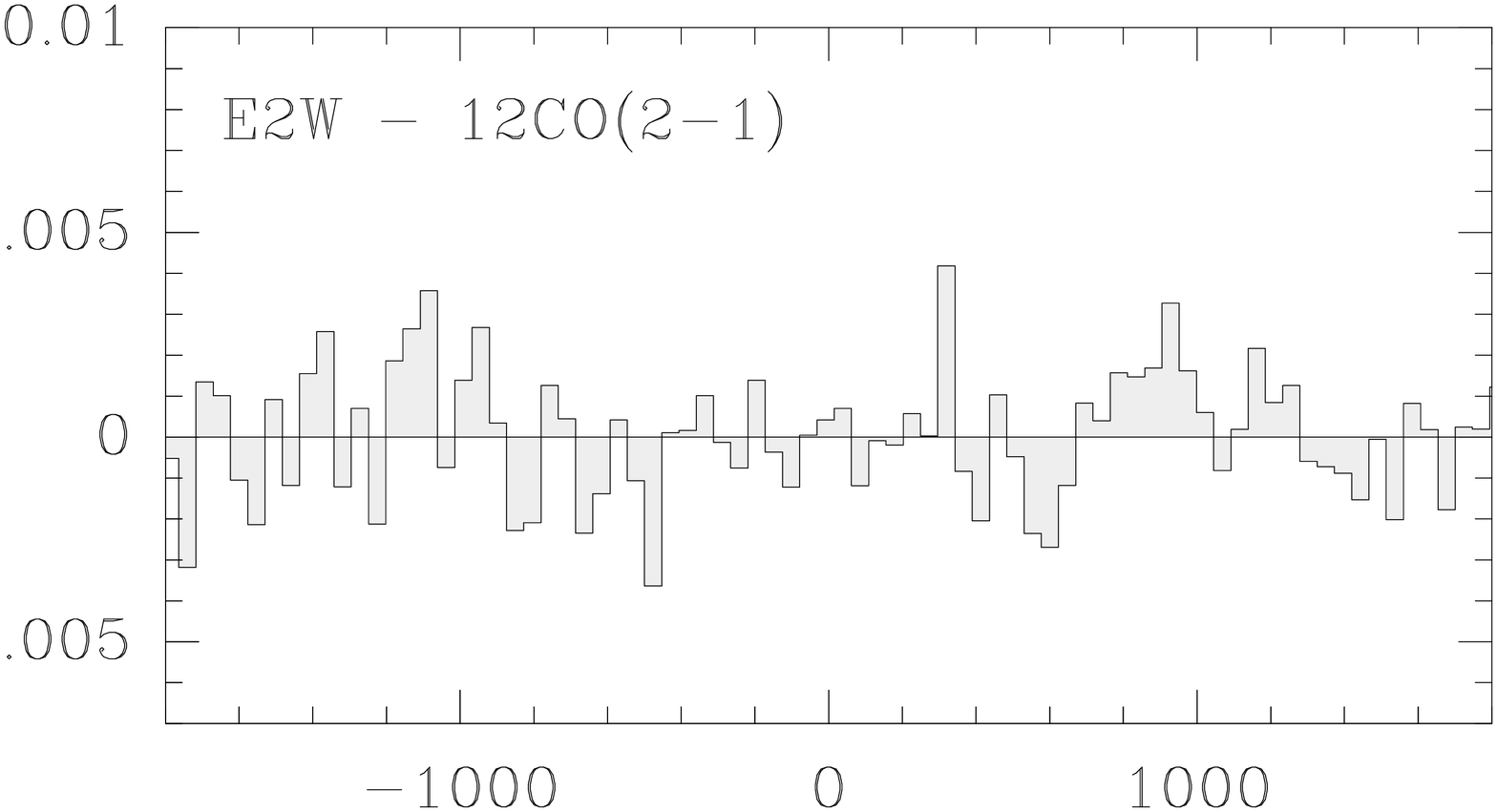}
} \\
\scalebox{0.93}{
\rotatebox{90}{~~~~~~~~T$_{mb}$ (K)}
\includegraphics[width=5.8cm]{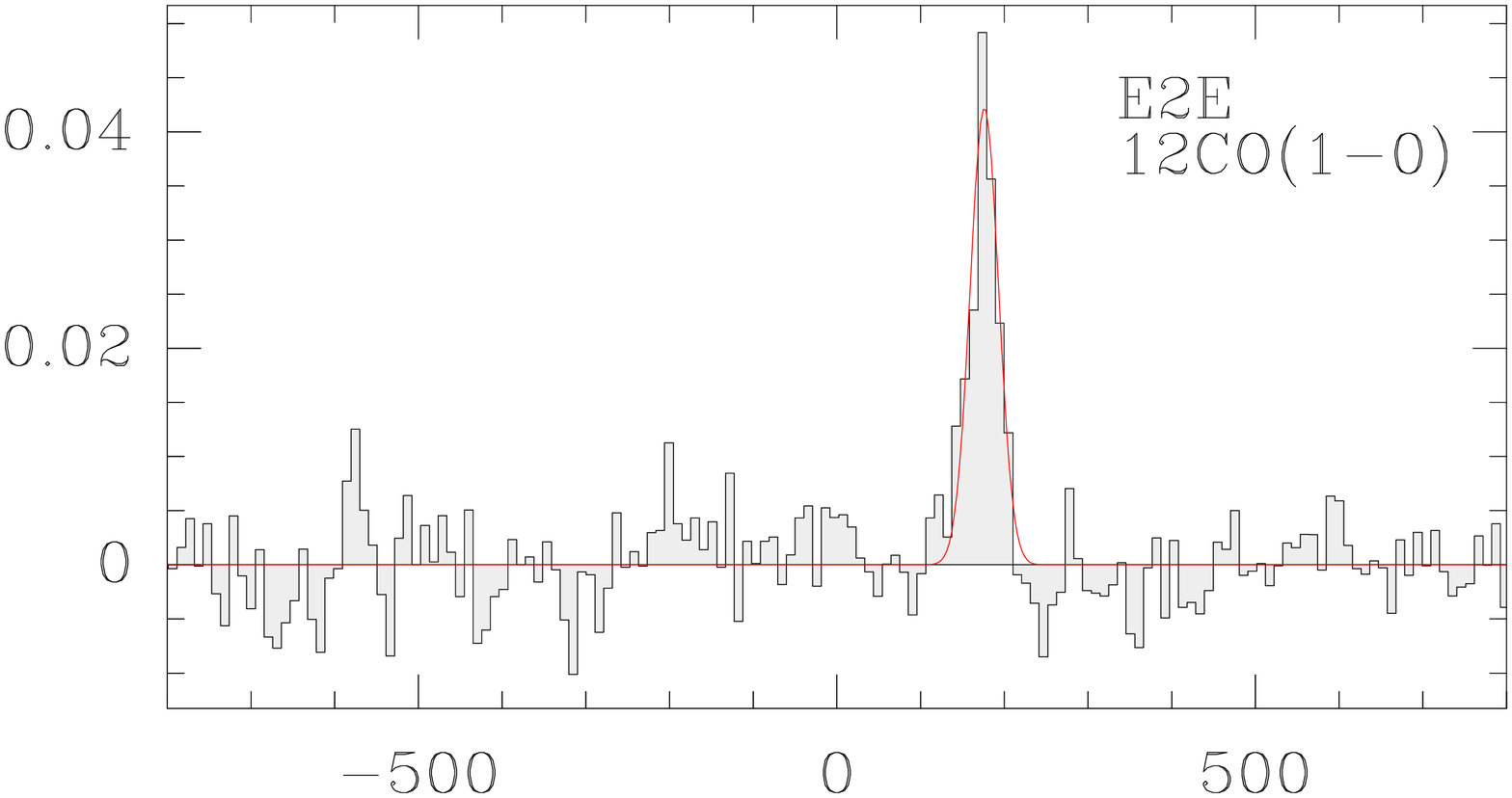}
\includegraphics[width=5.8cm]{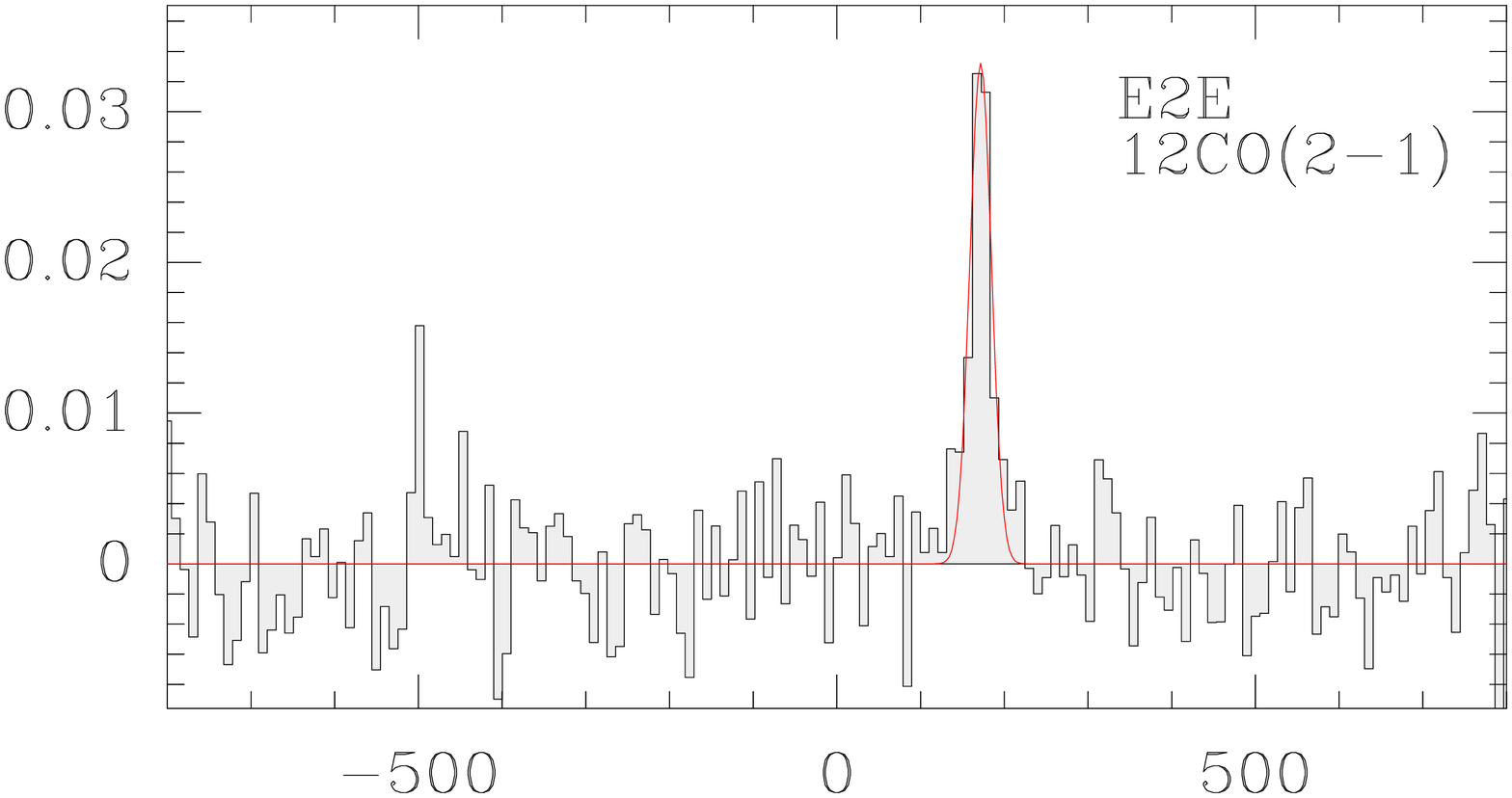}
\includegraphics[width=5.8cm]{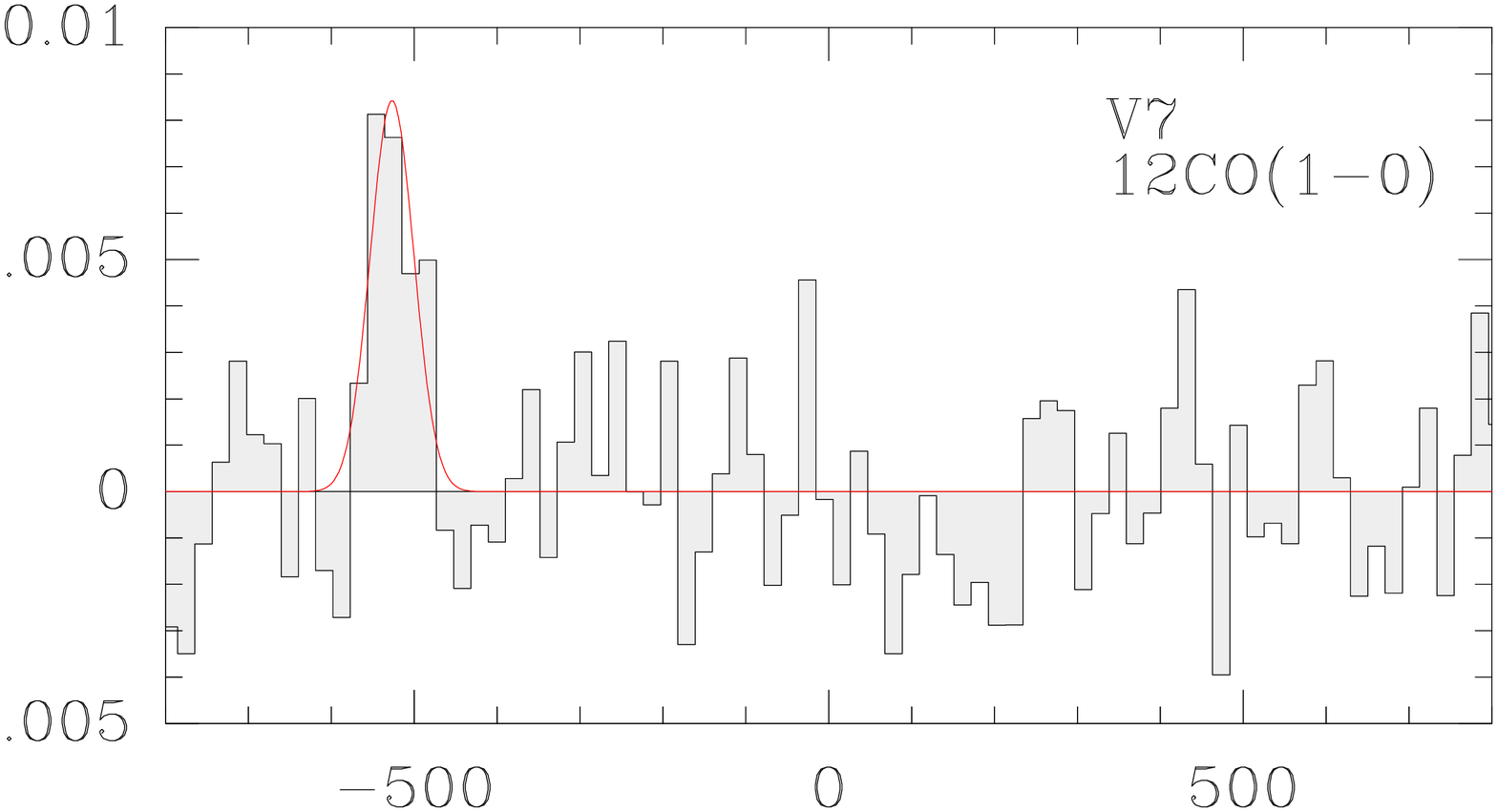}
} \\
\scalebox{0.93}{
\rotatebox{90}{~~~~~~~~T$_{mb}$ (K)}
\includegraphics[width=5.8cm]{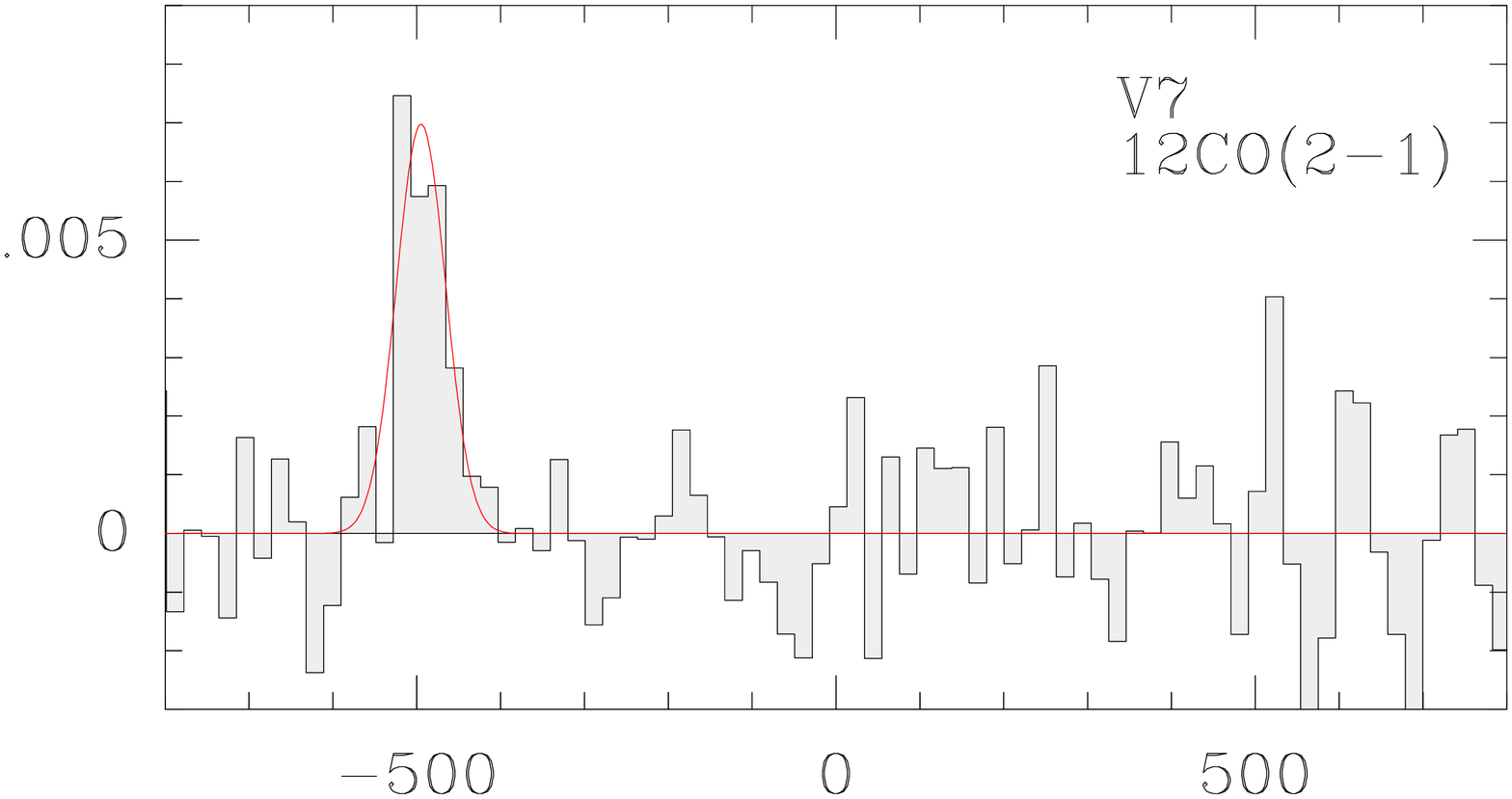}
\includegraphics[width=5.8cm]{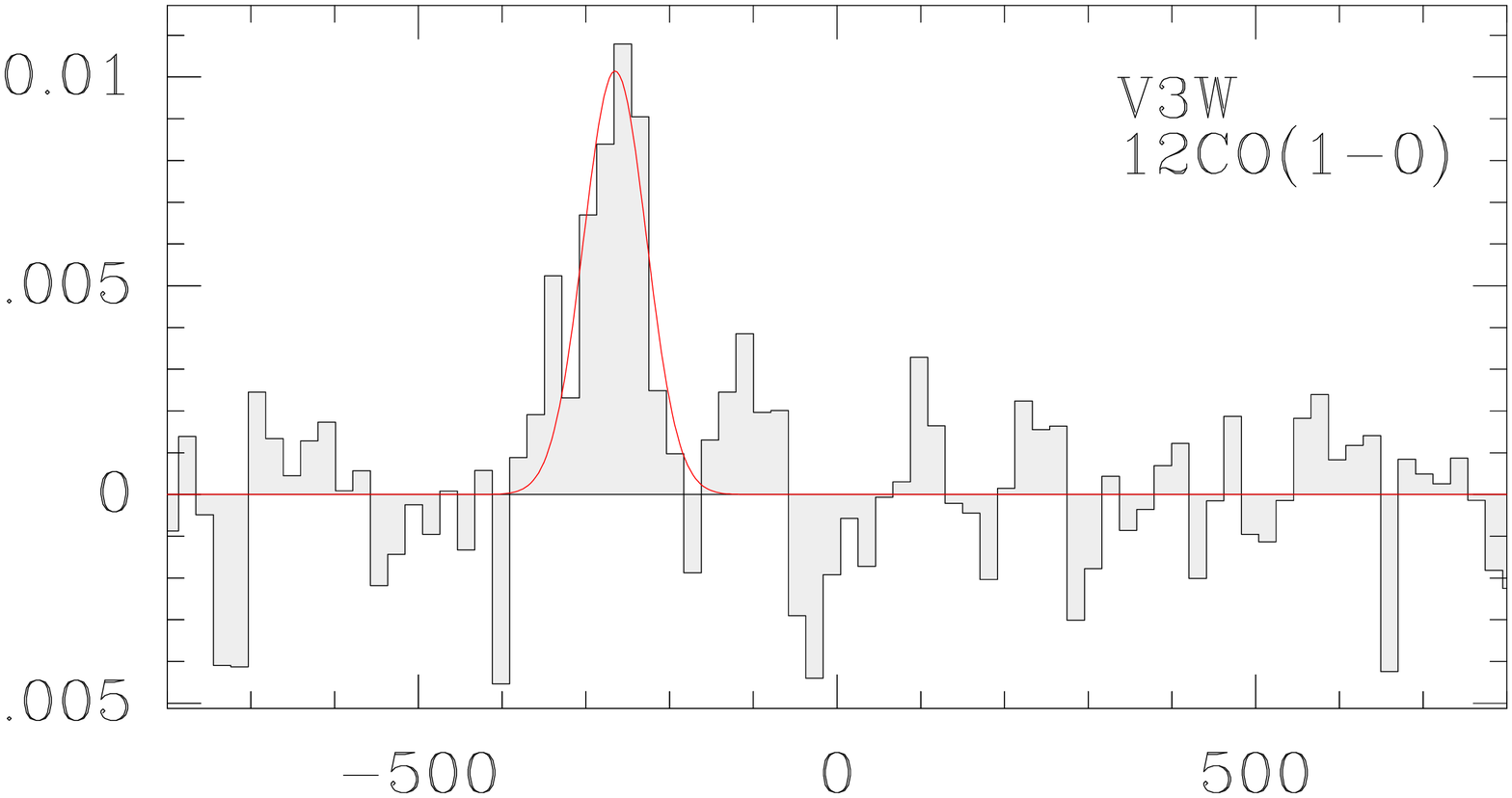}
\includegraphics[width=5.8cm]{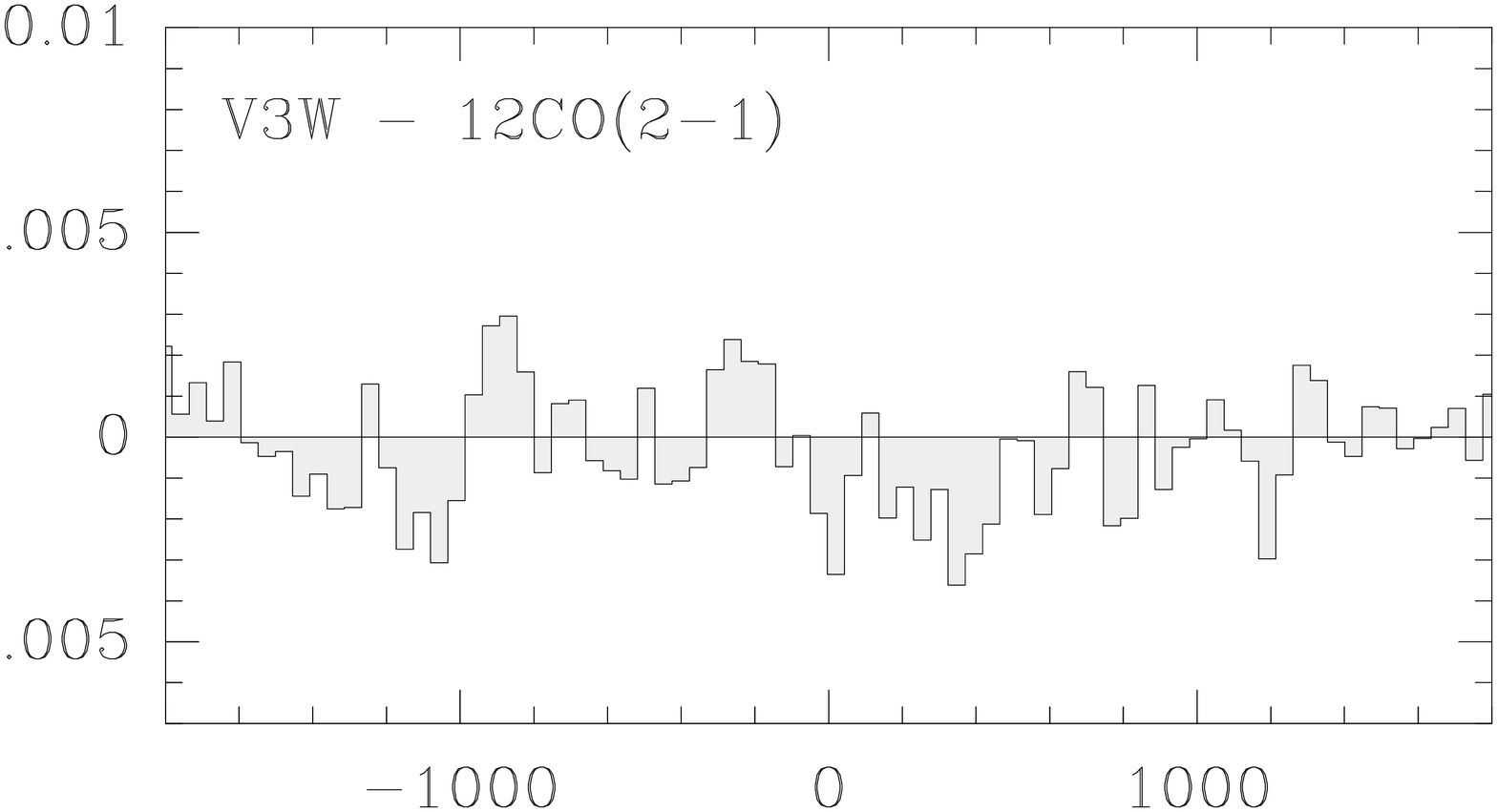}
} \\
\scalebox{0.93}{
\rotatebox{90}{~~~~~~~~T$_{mb}$ (K)}
\includegraphics[width=5.8cm]{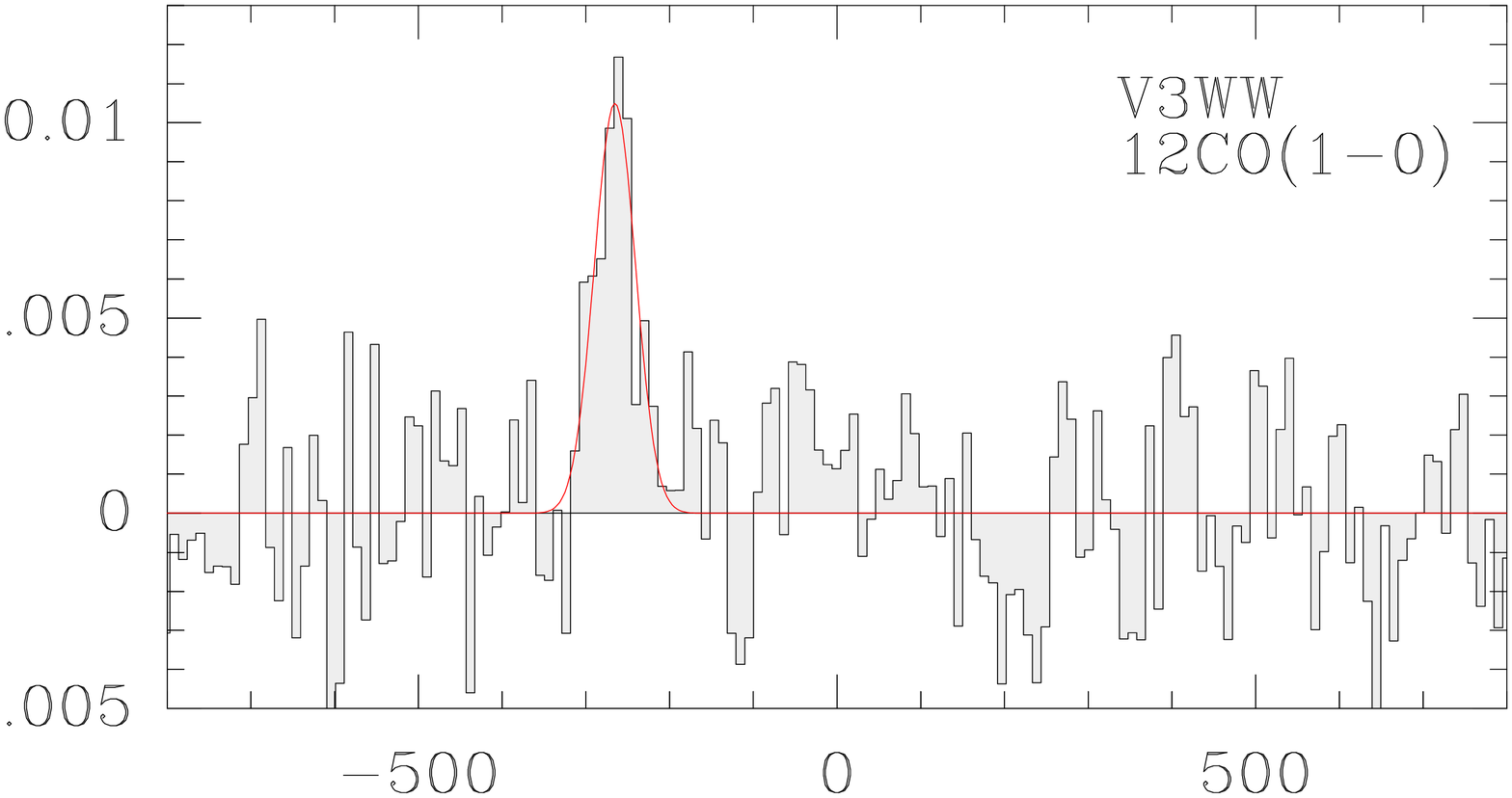}
\includegraphics[width=5.8cm]{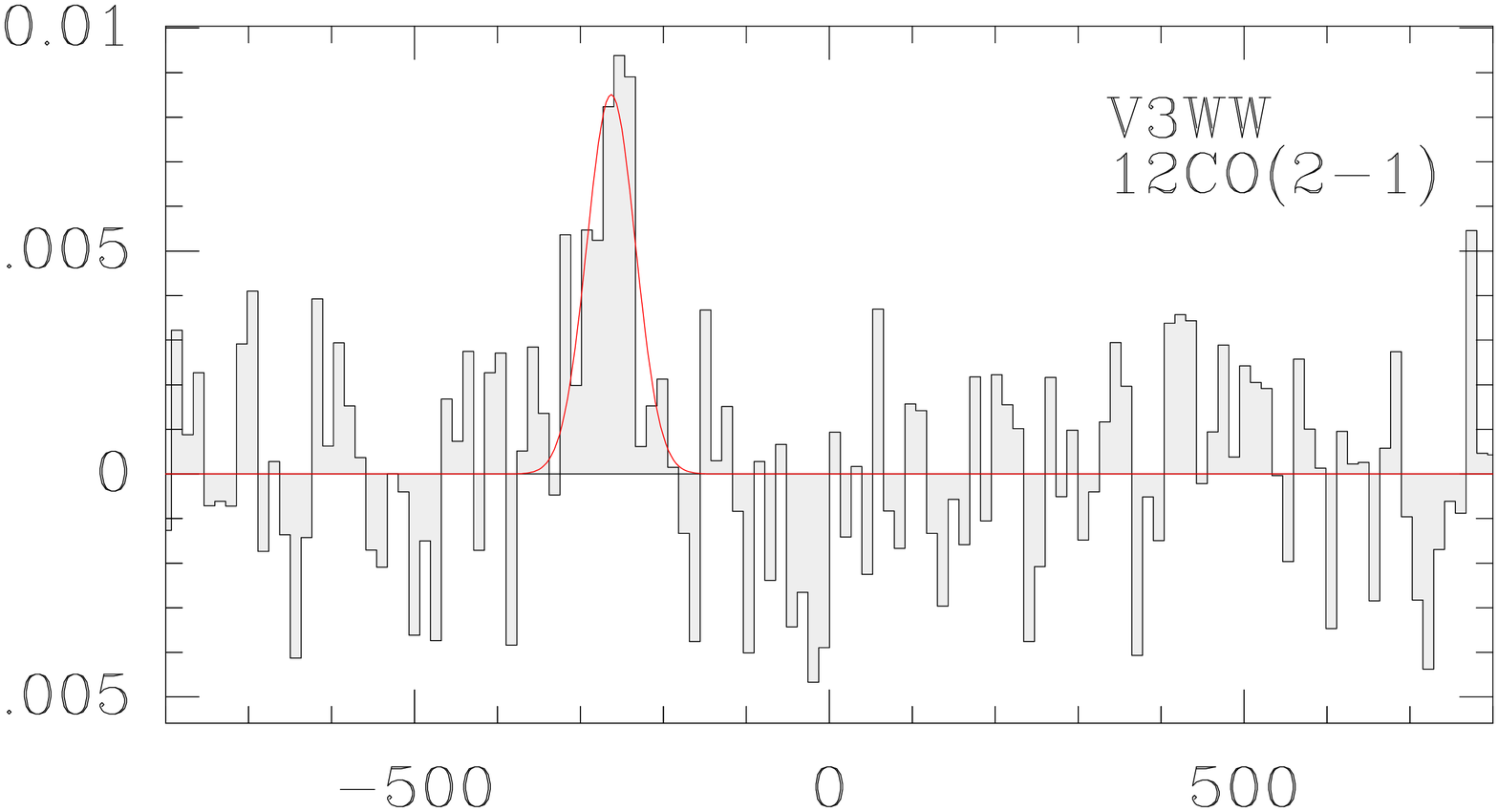}
\includegraphics[width=5.9cm]{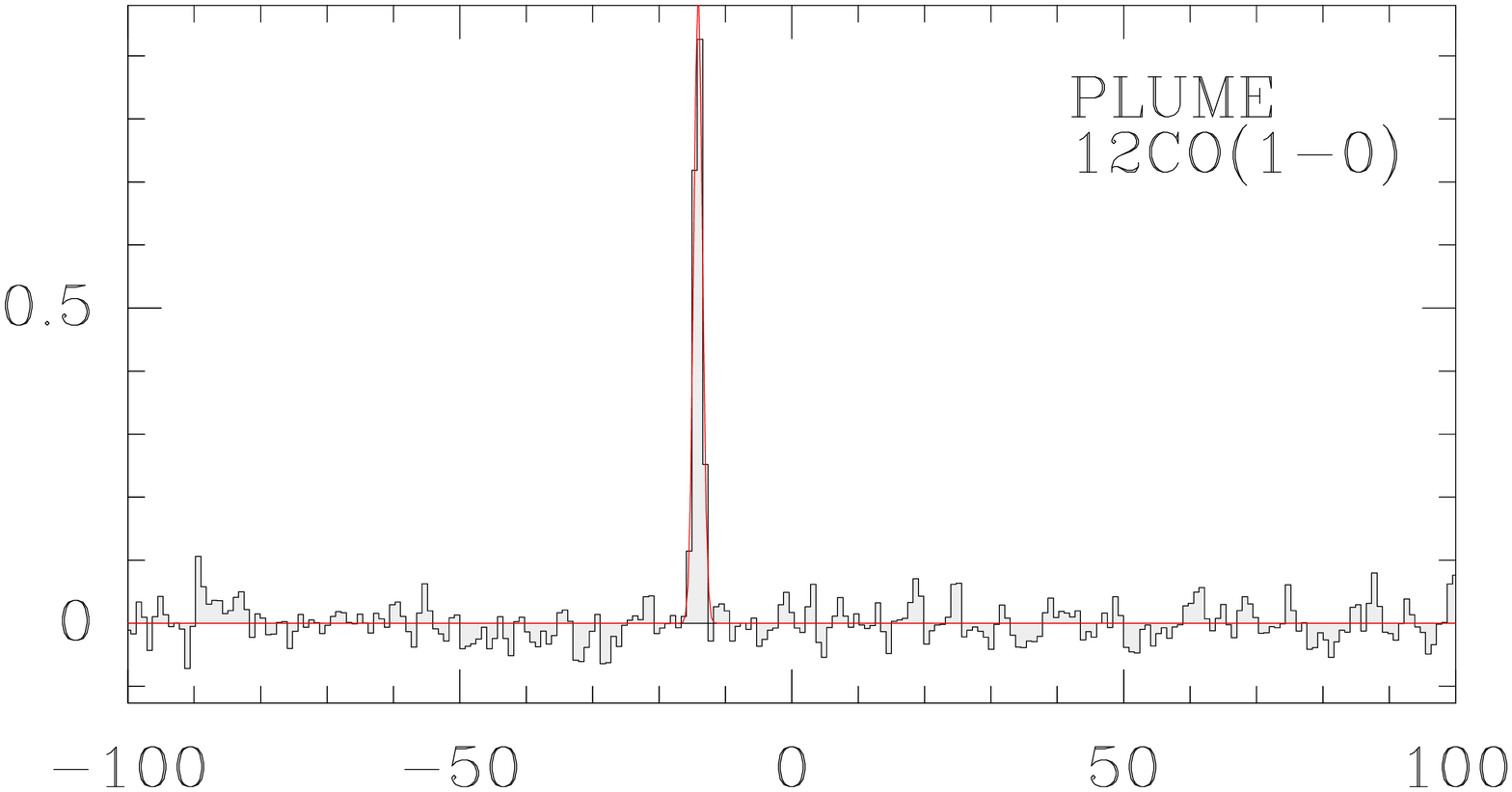}
} \\
\scalebox{0.93}{
\rotatebox{90}{~~~~~~~~T$_{mb}$ (K)}
\includegraphics[width=5.9cm]{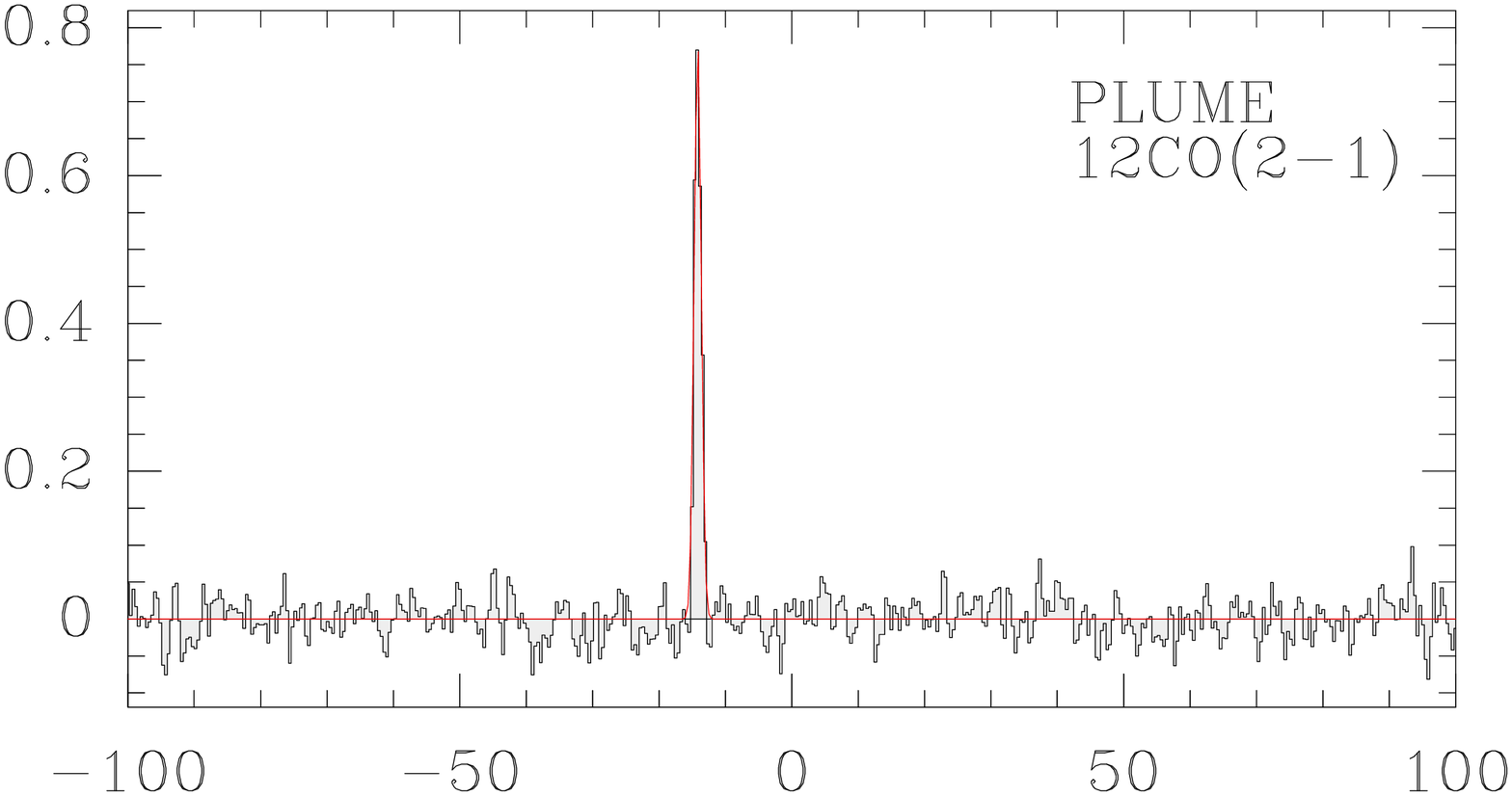}
\includegraphics[width=5.8cm]{f2_blank.eps}
\includegraphics[width=5.8cm]{f2_blank.eps}
}
\rotatebox{0}{~~~~~~~~~~~~~~~~~~~~~~~~~~~~~~~~~~~~~~~~~~~~~~~~~~~~~~~~~~~~~~~~~~~~~~Velocity (\kms )}
\caption{ Spectra of the plume and of the Virgo regions shown in Figure~\ref{fig:finding_chart} with \cooz\ line detections.
Gaussian fits to the CO line profiles (black) are overplotted in red.
}
\label{fig:lines}
\end{figure*}

\begin{table*}
\caption{\label{tab:data} \cooz\ and \coto\ line properties and derived gas masses in different Virgo regions.
}
\begin{centering}
\scalebox{0.9}{
\begin{tabular}{ p{1.6cm} c c c c c c c c c}
\hline
Position & Coordinates & Integration & Line &  $S(\rm CO)$\tablefootmark{a} & $V$\tablefootmark{b} & FWHM\tablefootmark{c} & $T_{\rm mb}$\tablefootmark{d} & resolution & $M_{\rm gas}$\tablefootmark{e} \\ 
 & (J2000) & time (min) & & (Jy\kms ) & (\kms ) & (\kms ) & (mK) & (\kms ) & (\msun ) \\
\hline\hline
NGC4438 & 12:27:45.60+13:00:32.0 &   15 & $^{12}$CO(1-0) & 226$\pm$6 & 72$\pm$6 &  395$\pm$12 & 108$\pm$14 & 10 & 5.7$\times$10$^8$ \\ 
NGC4438 & 					  &   15 & $^{12}$CO(2-1) & 208$\pm$8 & 61$\pm$7 &  362$\pm$14 & 108$\pm$15 & 13 & ... \\ 
A1              & 12:27:43.70+12:59:22.8 & 110 & $^{12}$CO(1-0) & 3.5$\pm$0.6 & -76$\pm$5 &  46$\pm$11 & 13.4$\pm$3 & 21 & 8.8$\times$10$^6$ \\ 
A1              &					  & 109 & $^{12}$CO(2-1) & ... &  ...  &  ...  & $<$5.1 & 47 & ... \\ 
A2              & 12:27:42.10+12:58:51.4 &   25 & $^{12}$CO(1-0) & 5.4$\pm$0.6 & -4$\pm$2 &  37$\pm$5 & 27$\pm$5  & 10 &  1.4$\times$10$^7$ \\ 
A2              & 					  &   25 & $^{12}$CO(2-1) & 3.8$\pm$0.6 & -12$\pm$3 &  29$\pm$6 & 24$\pm$5 & 10 & ... \\ 
E1              & 12:27:38.60+12:59:52.0 & 90 & $^{12}$CO(1-0) & 2.3$\pm$0.4 & 6$\pm$5 &  52$\pm$12 & 7.9$\pm$1.6 & 21 &  5.8$\times$10$^6$ \\ 
E1              & 					  & 90 & $^{12}$CO(2-1) & ... &  ...  &  ...  & $<$3.8 & 47 & ... \\  
E1W 	& 12:27:37.40+12:59:38.3 & 37 & $^{12}$CO(1-0) & ... &  ...  &  ...  & $<$5.4 & 47 &  $<$3.2$\times$10$^6$ \\ 
E1W 	& 					  & 37 & $^{12}$CO(2-1) & ... &  ...  &  ...  & $<$3.9 & 47 & ... \\ 
E2              & 12:27:44.00+13:03:00.0 & 50 & $^{12}$CO(1-0) & 8.8$\pm$0.7 & 144$\pm$2 &  42$\pm$5 & 39$\pm$3 & 10 & 2.2$\times$10$^7$ \\ 
E2              & 					  & 50 & $^{12}$CO(2-1) & 3.8$\pm$0.7 & 141$\pm$2 &  21$\pm$6 & 29$\pm$5 & 13 & ... \\ 
E2E           & 12:27:45.50+13:02:48.0 & 65  & $^{12}$CO(1-0) & 8.7$\pm$0.6 & 176$\pm$1 &  38$\pm$3 & 42 $\pm$4 & 10 & 2.2$\times$10$^7$ \\ 
E2E           & 					  & 65 & $^{12}$CO(2-1) & 5.3$\pm$0.6 & 172$\pm$2 &  28$\pm$5 & 33$\pm$4 & 10 & ... \\ 
E2W          & 12:27:42.10+13:03:15.0 & 80 & $^{12}$CO(1-0) & 2.6$\pm$0.6 & 157$\pm$10 &  84$\pm$22 & 6.1$\pm$2.1 & 21 & 6.5$\times$10$^6$ \\ 
E2W          & 					  & 80 & $^{12}$CO(2-1) & ... & ...  & ... & $<$4.8 & 47 & ... \\
V5 		& 12:27:31.60+13:01:38.0 & 42 & $^{12}$CO(1-0) & ... &  ...  &  ...  & $<$5.4 & 47 & $<$3.2$\times$10$^6$ \\ 
V5 		& 				 	  & 42 & $^{12}$CO(2-1) & ... &  ...  &  ...  & $<$3.8 & 47 & ... \\ 
V1 		& 12:26:56.20+12:59:40.0 & 52 & $^{12}$CO(1-0) & ... &  ...  &  ...  & $<$4.7 & 47 & $<$2.8$\times$10$^6$  \\ 
V1 		& 					  & 52 & $^{12}$CO(2-1) & ... &  ...  &  ...  & $<$6.4 & 47 & ... \\ 
V2		& 12:26:19.00+12:58:10.0 & 43 & $^{12}$CO(1-0) & ... &  ...  &  ...  & $<$4.7 & 47 & $<$2.8$\times$10$^6$ \\ 
V2		& 					  & 43 & $^{12}$CO(2-1) & ... &  ...  &  ...  & $<$4.1 & 47 & ... \\ 
V7 		&  12:26:15.90+12:58:49.0 & 50 & $^{12}$CO(1-0) & 2.8$\pm$0.5 & $-$527$\pm$6  & 59$\pm$12 & 8.4$\pm$1.9 & 21 & 7.1$\times$10$^6$  \\ 
V7 		& 					  & 45 & $^{12}$CO(2-1) & 2.6$\pm$0.4 & $-$495$\pm$6  & 66$\pm$13 & 7.0$\pm$1.4  & 21 & ... \\ 
V4 		& 12:25:47.50+13:02:44.0 &	23 & $^{12}$CO(1-0) & ... &  ...  &  ...  & $<$6.5 & 47 & $<$3.9$\times$10$^6$ \\ 
V4 		& 					  &	23 & $^{12}$CO(2-1) & ... &  ...  &  ...  & $<$4.4 & 47 & ... \\ 
V3 		& 12:26:17.00+12:54:54.0  & 37 & $^{12}$CO(1-0) & ... &  ...  &  ...  & $<$5.3 & 47 & $<$3.1$\times$10$^6$  \\ 
V3 		& 					  & 37 & $^{12}$CO(2-1) & ... &  ...  &  ...  & $<$5.6 & 47 & ... \\ 
V3W          & 12:26:14.90+12:54:54.0 & 129 &  $^{12}$CO(1-0) & 4.9$\pm$0.6 & -265$\pm$6 &  89$\pm$15 & 10$\pm$2 & 21 & 1.2$\times$10$^7$ \\ 
V3W          & 					  & 129 & $^{12}$CO(2-1) &  ... & ...  & ... & $<$4.4 & 47 & ... \\ 
V3WW      & 12:26:12.80+12:54:54.0 & 115 & $^{12}$CO(1-0) & 3.2$\pm$0.4 & -265$\pm$4 &  57$\pm$9 & 11$\pm$2.3 & 10 & 8.1$\times$10$^6$ \\ 
V3WW      & 					 & 115 & $^{12}$CO(2-1) & 3.1$\pm$0.5 & -263$\pm$6 &  68$\pm$16 & 8.5$\pm$2.2 & 13 & ... \\ 
V3NW 	& 12:26:15.50+12:55:22.0  & 40 & $^{12}$CO(1-0) & ... &  ...  &  ...  & $<$8.0 & 47 & $<$4.7$\times$10$^6$ \\ 
V3NW	& 					  & 40 & $^{12}$CO(2-1) & ... &  ...  &  ...  & $<$5.5 & 47 & ... \\ 
V6 		& 12:26:15.50+12:55.50.0  & 35 & $^{12}$CO(1-0) & ... &  ...  &  ...  & $<$7.5 & 47 & $<$4.4$\times$10$^6$ \\ 
V6 		& 					  & 35 & $^{12}$CO(2-1) & ... &  ...  &  ...  & $<$4.5 & 47 & ... \\ 
V8 		& 12:26:08.00+12:56:12.0  & 35 & $^{12}$CO(1-0) & ... &  ...  &  ...  & $<$7.3 & 47 & $<$4.3$\times$10$^6$ \\ 
V8 		& 					  & 35 & $^{12}$CO(2-1) & ... &  ...  &  ...  & $<$6.5 & 47 & ... \\ 
M86SW 	& 12:26:11.40+12:56:15.0 & 23 & $^{12}$CO(1-0) & ... &  ...  &  ...  & $<$12 & 47 & $<$7.1$\times$10$^6$ \\ 
M86SW 	& 					  & 23 & $^{12}$CO(2-1) & ... &  ...  &  ...  & $<$12 & 47 & ... \\
PLUME    & 12:27:30.00+13:12:12.0 &   10 & $^{12}$CO(1-0) & 8.1$\pm$0.2 & -14  &  1.2 & 1018$\pm$27 & 0.8 & ... \\ 
PLUME    & 					 &   10 & $^{12}$CO(2-1) & 5.2$\pm$0.1 & -14  &  1.2 & 782$\pm$27 & 0.4 & ... \\ 
\hline
\end{tabular}}
\end{centering}
\tablefoottext{a}{Fluxes obtained from the Gaussian fitting to the line profiles, and multiplied by a 
temperature-to-flux conversion factor of 5.0 Jy/K that is appropriate for the 30m telescope at all observed frequencies.} 
\tablefoottext{b}{All velocities are heliocentric.} 
\tablefoottext{c}{Resolution-corrected line width.} 
\tablefoottext{d}{All limits are computed at 3$\sigma$ levels.}
\tablefoottext{e}{A standard CO-to-\htwo\ Galactic value of 4.6 \msun\, (K \kms pc$^2$)$^{-1}$ was used for the 
mass computation. The 3$\sigma$ limits were computed for a width of 47 \kms . This is close to the median \cooz\ 
width for all regions but the plume and NGC4438.} 
\end{table*}

\begin{figure*}
\begin{center}
\includegraphics[width=18.8cm]{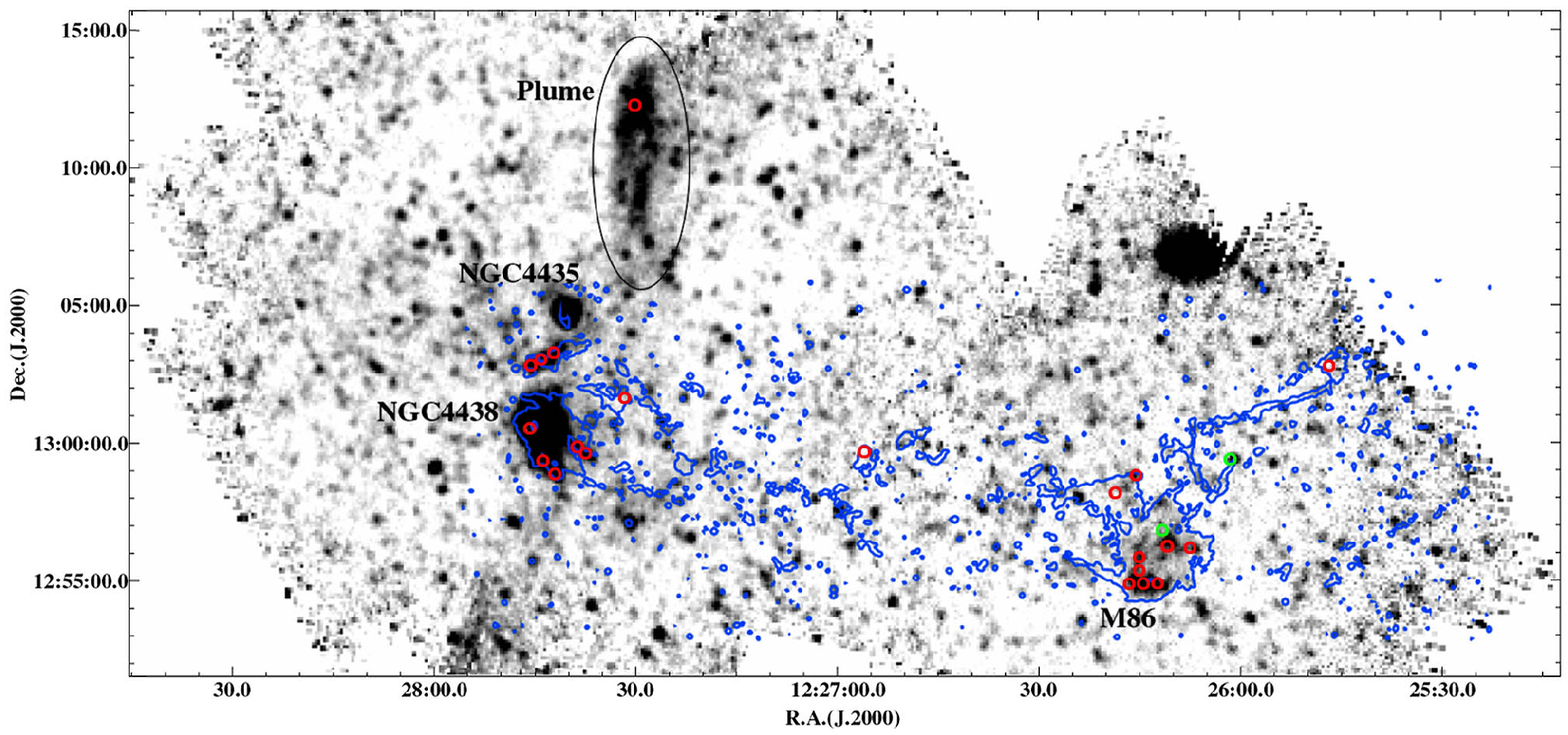} 
\caption{ Our CO observation positions (red circles) and positions of previous CO observations (\citealt{wiklind95}; \citealt{braine97}; green circles),
over \her\ 250 \um\ SPIRE imaging data (adapted from \citealt{cortese10b}).
The \ha\ intensity contours are overplotted in blue.
}
\label{fig:finding_chart_spire}
\end{center}
\end{figure*}

\section{Observations and data analysis}
\label{sec:data}

The observations were carried out with the 30m telescope at Pico Veleta, Spain. We observed twenty \ha -emitting 
regions in or near NGC4438 and M86, and along the tidal bridge that connects them (Figure~\ref{fig:finding_chart}), 
to test whether they contain detectable amounts of CO. In addition to these positions, we also observed clouds in the 
diffuse, X-ray emitting plume northwest of NGC4438 and NGC4435, which is most likely attributed to Galactic cirrus 
\citep{cortese10a}.

We simultaneously observed the $^{12}$CO J=1$-$0 and J=2$-$1 lines with the EMIR receivers. The receivers were 
tuned to the frequencies of the two transitions at 115.271 GHz and 230.538 GHz, respectively, shifted to the velocity of 
the \ha\ gas in each position. All of the 4MHz, WILMA, and VESPA backends, with corresponding resolutions of 4 MHz, 
2 MHz, and 0.3 MHz, were simultaneously attached to allow consistency checks, with the WILMA data being primarily 
used in this work. VESPA data are only shown for the plume. The observations were performed in wobbler-switching 
mode with a throw of 2\farcm40. The telescope pointing was monitored by observing a bright radio source every two hours, 
indicating a pointing accuracy of 2\arcsec\ to 3\arcsec. The system temperature varied from 270 to 500\,K at 3\,mm.
We typically spent 0.25-2 hours on source, which led to 1$\sigma$ noise levels of 2 mK to 10 mK at the frequency 
of the \cooz\ transition (Table~\ref{tab:data}). 

The data were reduced with the CLASS software of IRAM. We averaged the spectra of each region after removing 
bad channels and subtracting linear baselines, dropping scans with baseline fluctuations. The resulting spectra, 
binned to 10-21  \kms\ channels, are shown in Figure~\ref{fig:lines}. For the plume, the high-resolution VESPA 
spectra are unbinned. A Gaussian function was fitted to each line for the derivation of its width, peak position,  
temperature $T$, and flux $S$, which are all shown in Table~\ref{tab:data}. Assuming that the observed clouds are 
unresolved, $T$ corresponds to the main beam temperature. It was converted to a flux using a temperature-to-flux 
conversion factor of 5.0 Jy/K (appropriate for both the 1 and 3 mm observations). 

The cold \htwo\ gas masses, which are also presented in Table~\ref{tab:data}, were computed as
\begin{equation}
\label{eq:plane}
M_{\rm H_2} = \alpha\, \frac{ 23.5\, I_{\rm {CO}}\, \Omega_B\, {D_L^2}}{(1+z)^3} \msun ,
\end{equation}
where $I_{\rm {CO}}$ is the 115GHz line intensity in K \kms , $\Omega_B$ is the main beam area in arcsec$^2$ (for 
unresolved clouds), $D_L$ is the luminosity distance in Mpc, and $\alpha$ is the CO luminosity to \htwo\ mass conversion 
factor \citep{solomon97}. We used a standard Galactic conversion factor of 4.6\msun\,(K\,\kms\,pc$^2$)$^{-1}$ 
\citep{bloemen86,downes93}, and a telescope beam of 22\arcsec\ (i.e., 1.8\,kpc) at 3\,mm. The 3$\sigma$ mass limits for 
non-detections were computed using a resolution (and width) of 47 \kms , which is characteristic of the clouds under 
examination (see Table~\ref{tab:data}).

\section{Results: CO-emitting vs non-emitting regions}
\label{sec:results}

Narrow \cooz\ and \coto\ emission is found in the plume clouds northwest of NGC4438 and NGC4435, at 12:27:30$+$13:12:12. 
Both line widths are of only 1.2 \kms , confirming that the emission originates from Galactic cirrus clouds. This possibility was  
suggested by \citet{cortese10a} based on the \cooz\ line width at a nearby position, 12:27:30$+$13:12:29.  CO lines were detected 
in another 10 of the 20 regions shown in Figure~\ref{fig:finding_chart}, but their kinematics were atypical of cirrus clouds, with line
width(s) exceeding 30 \kms\  in all cases. A cross examination of the high-resolution VESPA data confirms this finding. 

Several regions in Virgo, at $\sim$10 kpc from the $B$-band center of M86 (at 12:26:12+12:56:49) had a CO detection. The region 
V7, which is located 11 kpc northeast of M86 and in the northwest tip of the local \ha\ filamentary structure (Fig.~\ref{fig:finding_chart}),  
contains 7.1$\times$10$^6$\msun\ of gas that is moving at $-$527\kms . This is very close to the measured \ha\ recession velocity, $-$500\kms\ 
\citep{kenney08}. This line-of-sight velocity precludes a clear answer to the question of whether the clouds are bound to M86. The escape velocity 
at a distance of 10\,kpc from the center of M86 is high, 470\kms , because its bulge mass equals 2.6$\times$10$^{11}$\msun . The latter is given 
in units of 10$^{10}$\msun\ by $c_2 {\sigma _{100}}^2$\reff , where $c_2$ is a unitless coefficient depending on the matter distribution,
$\sigma _{100}$ is the stellar velocity dispersion in units of 100\kms , and \reff\ is the effective radius in kpc. The value of $c_2$ for giant 
ellipticals  is 1 \citep{bender92}. For M86, $\sigma _{100}$ is 2.2 \citep{smith00} and \reff\ is 5.5 kpc \citep{gavazzi05} in the $B$ band. 

\begin{figure}
\begin{center}
\includegraphics[width=8cm]{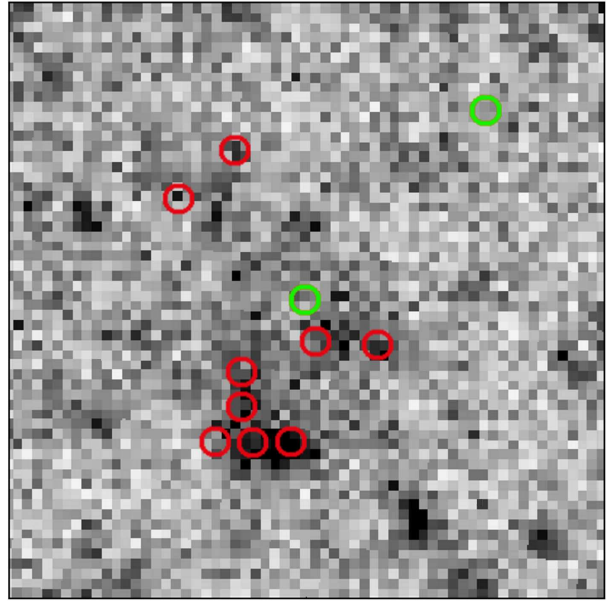} 
\caption{Our observations plotted over a 4\arcmin$\times$4\arcmin\ zoom around M86 at 350\um . The far-infrared image is taken from 
the first public data release of the \her\ Virgo Cluster Survey (HeVICS), which was initially presented in \citet{davies10}. The CO positions 
are color-coded as in Figure~\ref{fig:finding_chart_spire}.}
\label{fig:map350}
\end{center}
\end{figure}

CO gas was also found in the V3W and V3WW regions, along an \ha\ stream that results in the center of M86, at 9 to 10 kpc away from
it (Figure~\ref{fig:finding_chart}). The CO emission coincides with the H\,I emission peak of M86 \citep{li01,kenney08}.  The recession velocity 
of the CO, $-$265\kms , agrees well with that measured from the H\,I,  $-$240\kms\ \citep{bregman90}, and from the Mg stellar absorption 
features in optical wavelengths, $-$244\kms\ \citep{smith00}. The \cooz\ FWHM in these regions, 50-90\kms , is comparable with the narrow 
H\,I component FHWM, 60\kms\ \citep{bregman90}. The combined \htwo\ mass for both V3W and V3WW regions is 2.0$\times$10$^7$\msun . 
If stars are forming out of this gas, and if the galaxy-integrated SFR calibration of \citet{gao04} is applicable for these clouds, then the SFR will 
be 0.03\msun\,yr$^{-1}$. Likewise, the combination of the modified black body fit to SPIRE data \citep[][ Figs.~\ref{fig:finding_chart_spire} and 
\ref{fig:map350}]{cortese10b} of the same regions that was performed by \citet{gomez10} and the \citet{kennicutt98} relation lead to a dust-based 
SFR estimate that is also on the order of 0.01\msun\,yr$^{-1}$. 

No molecular gas was found at their neighboring V3, V3NW, V6, V8, and M86SW regions. The latter is along the same \ha\ stream connecting 
V3W and V3WW to the very center of M86. For T$<$12mK and for a line width of 47\kms , the gas mass in M86SW is $<$7$\times$10$^6$\msun . 
Molecular gas has not been detected at the center of M86 either \citep{wiklind95,braine97}; if it exists, its mass must be below 5$\times$10$^6$\msun\ 
for the assumptions in this work. 

No gas was found either at V1, V2, and V5 of the tidal bridge between NGC4438 and M86, or at V4, which is located in the figure-of-eight shaped 
\ha\ loop northwest of M86\footnote{This loop belongs to Virgo's ICM and part of it could be bound to M86, because its \ha\ recession velocity 
ranges from $-$230 to $-$450 \kms .}. If gas does exist in any of these regions, its mass is below 4$\times$10$^6$\msun . Previous observations by 
\citet{wiklind95} at another location of this loop, 12:26:02$+$12:59:21, also led to a non-detection, with an upper limit of 7.9$\times$10$^6$\msun . 

Most of the CO gas is detected in or around NGC4438. At the center of NGC4438, the telescope beam encompasses 6$\times$$10^8$\msun\ 
of \htwo\ gas that has a recession velocity identical to that of the \ha . The \cooz\ line profile also agrees well with that found from previous 
observations by \citet{vollmer05}. Both the \cooz\ and the \coto\ profiles indicate that the gas is in a disk whose rotational velocity reaches 
200\kms\ within the inner 4 kpc of NGC4438\footnote{
 \hbox{The differences in the observed \cooz\ and\,\coto } line properties of other regions, such as E2W, E2, and E2E, can be accounted for by
differences in the gas density and excitation properties, or differences in the sensitivity and the beam size of the observations at the two frequencies,
if some filaments are resolved.}.
New observations are presented here for the gas clouds at E1, and in the regions E2, E2E, and E2W at $\sim$10 kpc north of the nucleus. 
We specifically observed parts of the \ha\ filament that is oriented perpendicularly to the tidal tail northeast of NGC4438, which was
previously observed by \citet{vollmer05}. The nearest positions of our observations and those of \citet{vollmer05} are one beam (22\arcsec ) 
away. The clouds in the E2 complexes are moving at $+$70 to $+$100\kms\ from systemic velocity along the line of sight. Their total \htwo\ mass 
is 5.1$\times$10$^7$\msun . Their counterparts in the south of the nucleus are named A1 and A2. They  contain 2.3$\times$10$^7$\msun\ of 
\htwo\ gas, and they are moving at $-$80 to $-$150 \kms\ from systemic velocity. For both the northern and southern clouds, the observed 
velocities are consistent with gas that is either part of a rotating disk, or experiencing tidal and ram pressure stripping \citep{vollmer05}. 
If all of these A and E2 cloud complexes are stripped, a considerable (14\%) fraction of the gas in the center of NGC4438 (see 
Table~\ref{tab:data}) is potentially lost to the ICM and to neighboring galaxies.

\section{Discussion: the origin of the molecular gas near M86}
\label{sec:discussion}
 
\citet{li01} attributed the origin of the H\,I\ emission, which peaks near CO-emitting regions, to gas that cooled down from a
compressed, shocked, and heated medium. In cooling flows, however, the ICM gas is depleted of its dust due to sputtering by 
X-rays.  Dust sputtering leads to gas-to-dust ratios that exceed their typical values for local galaxies \citep[i.e. several 
hundreds;][]{young91,wiklind95} in timescales of only Myr in 10$^6$-10$^7$K environments \citep{clemens10}. On the other hand, 
the presence of dust in all M86 regions where CO was detected \citep[Figs.~\ref{fig:finding_chart_spire} and \ref{fig:map350};][]{cortese10b,
gomez10} indicates that the gas and the dust were commonly accumulated. Their relative mass ratio indeed favors the capture 
from an external galaxy rather than the formation from a cooling ICM. For the inner 2\arcmin\ of M86, the total 15-20\,K dust mass 
is (2-5)$\times$10$^6$\msun\ \citep{gomez10}. If the molecular gas distribution follows the dust distribution, most of the \htwo\ gas 
in the inner 2\arcmin\ will be accounted for by the observations at V6, V3NW, V3W, V3WW, V8, M86SW, and at the center of M86. 
The total \htwo\ mass will then be in the range (2-5)$\times$10$^7$\msun. Furthermore assuming that the center of M86 comprises 
1.1$\times$10$^7$\msun\ of atomic gas based on its \ha +\nii\ luminosity \citep{kenney08}, and another 5.6 $\times$10$^7$\msun\ 
based on its H\,I observations \citep{bregman90}, we deduct that its gas-to-dust ratio will be in the range 20-60. The range 
can be broader when considering the different beam sizes of the observations. For comparison, \citet{cortese10b} give a 10-20\,K 
dust mass of 2$\times$10$^6$-2$\times$10$^7$\msun , leading to a gas-to-dust ratio of 30-300 (including helium) for a 
dust knot 5 kpc northwest of NGC4438. 


\begin{figure*}
\begin{center}
\includegraphics[width=16cm]{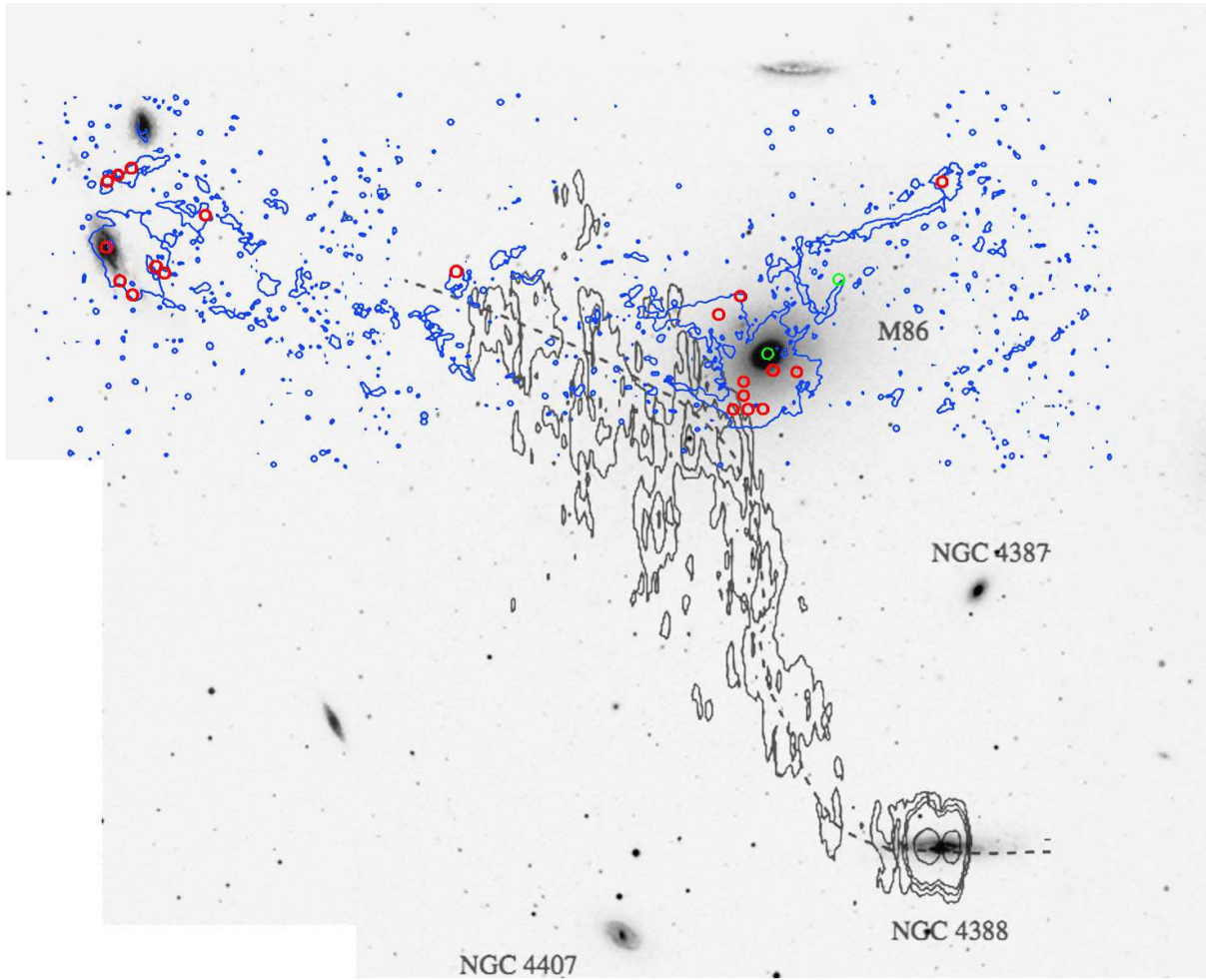} 
\caption{ CO observation positions and \ha\ intensity contours (as in Figure~\ref{fig:finding_chart_spire}) and H\,I intensity contours 
(black; \citealt{oosterloo05}) plotted over a DSS B-band image.}
\label{fig:ha_hi}
\end{center}
\end{figure*}
 
With its \ha -emitting trail, NGC4438 is the best candidate for the external capture of the gas and the dust in M86. While the trail started 
elongating as the two galaxies were receding from each other, most of the gas and the dust was likely captured when the 
galaxies where near the pericenter of their encounter. If gas was transferred from 
NGC4438 to M86 in ionized or neutral state through dynamical instabilities or ISM-ISM stripping, CO could be detected in M86 if it 
reformed in situ. Molecular formation is possible to occur in situ in the V3W and V3WW complexes. The column density threshold for the 
efficient conversion of H\,I to \htwo\ is in the range 3$\times$10$^{20}$-10$^{21}$ cm$^{-2}$ \citep{schaye04}. The \hi -deduced 
column density in these regions is close to the $1.6\times 10^{20}$cm$^{-2}$ contour of \citet{kenney08}. Likewise, the CO-based 
\htwo\ column density is 2.9$\times$10$^{20}$ and 1.9$\times$10$^{20}$cm$^{-2}$ for V3W and V3WW, respectively, for the adopted 
CO-to-$\htwo$ Galactic conversion factor. The H\,I column density requirement would be satisfied if the H\,I-to-\htwo\ mass ratio 
was $\gtrsim$3, which is comparable to that of both moderately stripped spirals and ellipticals \citep{kenney89,wiklind95}. The present 
computations assume that the gas emission is smeared out across each telescope's beam. When taking into account the clumpy distribution 
of the clouds, the column density requirement can be easily met. For this scenario to hold,  the CO needs to be detected in regions where 
dust is also present, because the dust particles act like catalysts for the H\,I atoms to recombine upon. This condition is also satisfied.

Contrarily to the V3W(W) complexes, the bulk of the molecules could not have formed in situ in V7, where the \hi -deduced column density  
is outside the last contour of \citet{kenney08}, i.e., below 10$^{19}$cm$^{-2}$. The detection of CO there provides evidence for the survival 
and transport of self-shielded molecular clouds in the hot ICM of Virgo. The molecules could have started forming, e.g., near the \hi\ peak 
south of M86 and then rotated around it in a half-circular orbit of $\sim$30 kpc. If the clouds were moving at 300\kms\ (i.e. close to the V3W 
and V7 difference in recession velocity), their travel time would be 100 Myr. Because the formation of new stars from \htwo\ in isolated clouds 
takes place in timescales of 10$^5$-10$^7$ yr \citep[e.g.,][]{krumholtz11}, this result implies that the collapse, the fragmentation, and the 
reformation of the molecular clouds from gas in their vicinity is a continuous process inside filaments. 
 
Alternatively, gas clumps could have been transferred from NGC4438 to M86 directly in a dense, molecular state. The molecular gas has then 
survived near M86 for a time comparable to the travel time of NGC4438 from the pericenter to its present location. The orbital simulations of 
\citet{vollmer09} also suggest that NGC4438 has been moving through the hot X-ray gas for over only 100 Myr after its encounter with M86.
This timescale is insufficient for the various energy/momentum transport processes that act on the molecular clouds to fully destroy them. These 
processes include the heat conduction from the surrounding medium, the viscous flow stripping, and the Kelvin-Helmholtz instabilities, with 
the heat conduction leading to the highest mass loss rates in this environment \citep{nulsen82}. In the ICM, the mean free path of the electrons is 
large compared to the radii of the molecular clouds, leading to saturated heat conduction and long evaporation timescales \citep{cowie77}. For 
a cloud with radius of 10\,pc and density $n$ of 10$^3$\,cm$^{-3}$ that is embedded in a 10$^7$K medium with density of 10$^{-4}$\,cm$^{-3}$, 
the evaporation time is 3 Gyr. This time can increase when magnetic fields and radiative cooling are taken into account \citep{cowie77,mckee77}, 
because cooling is very efficient above 10$^4$K \citep{dalgarno72}. It effectively dominates over heating for clouds that are stripped outside 
of galaxies and that are not sufficiently heated by stellar ultraviolet radiation \citep{vollmer01}. The atomic gas will then cool down, condense, 
and reform \htwo\ within $\sim$10$^9$/$n$ yr\,cm$^{-3}$ \citep{hollenbach97}. It is therefore plausible that the molecular gas observed in M86 
was directly stripped from NGC4438, even though it was not detected in our observations of, e.g., the V1 region. This might occur if there is no
strong source of gas excitation, such as star formation, shocks, cosmic rays \citep{ferland08}, or diffuse intracluster light \citep{mihos05}.


\begin{figure}
\begin{center}
\rotatebox{90}{~~~~~~~~~~~~T$_{mb}$ (K)}
\includegraphics[width=8.4cm]{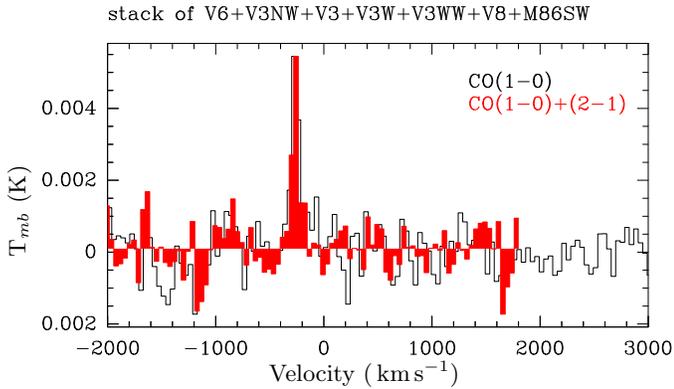}
\rotatebox{0}{~~~~~~Velocity (\kms )}
\caption{Stacked spectra of all positions south of M86. No gas is detected at any velocity other than the galaxy's recession velocity.}
\label{fig:stack}
\end{center}
\end{figure}

An entirely different scenario that remains to be tested is whether the gas in M86 could originate from NGC4388, a spiral 
galaxy at $\sim$100kpc in projection on its south, with an estimated \htwo\ reservoir of 10$^8$-10$^9$ \msun\ 
\citep{kenney89,papadopoulos98}. NGC4388  could be interacting with the M86-NGC4438 complex, if all galaxies 
were at a sufficiently close distance  \citep[e.g.,][]{willick97}. The interaction scenario was considered after the detection of 
a long ($>$120 kpc) trail of neutral gas starting from NGC4388 and heading in the direction of M86 and NGC4438
\citep[][Fig.~\ref{fig:ha_hi}]{oosterloo05}, and by the strong gas stripping in this system \citep[e.g.,][]{kenney89,vollmer09}. 
Very Large Array data indicated that NGC4388 has a radio jet that is driven by an active nucleus \citep{colbert96,irwin00} 
and that could be adding to the gas expulsion mechanisms. A massive outflow has been observed in its \ha\ and \nii\  images 
\citep{veilleux99,kenney08}, with an \ha\ velocity ranging between 2000\kms\ and 2400\kms\ \citep{yoshida04}. The velocity 
of the neutral, H\,I-emitting gas along the trail is likewise between 2000-2700\kms\ \citep{cayatte90,oosterloo05,chung09}.
The projected H\,I detection limits are close to the \ha\ detection limits in the vicinity of M86 (Fig.~\ref{fig:ha_hi}). However, 
the non-detection of CO moving at any velocity between $-$250\kms\ and 2700\kms\ in the south of M86 (either in individual or
stacked spectra; Fig.~\ref{fig:stack}) renders this scenario improbable. A similar conclusion was drawn by \citet{oosterloo05} 
based on the H\,I gas kinematics. Additional (and deeper) observations would be needed to test the viability of this scenario.

\section{Summary}
\label{sec:conclusions}
Using IRAM's 30m telescope, we investigated the presence of molecular gas in Virgo's hot ICM, and specifically in the vicinity of
M86, NGC4438, and in the 120 kpc-long tidal trail that connects them. We focused on \ha\ emission peaks, which could originate 
from the interface of dense clouds with self-shielded CO molecules and the X-ray emitting ICM, because ram pressure stripping 
has dispersed most of the diffuse ionized gas. We observed 21 regions  in the \cooz\ and \coto\ transitions and we draw the following 
main conclusions.
\begin{itemize}
\item 
Narrow CO emission, with line widths of only 1.2 \kms , was detected at 12:27:30$+$13:12:12, northwest of NGC4438 and NGC4435.
It is tracing Galactic cirrus clouds. In no other position were the CO line widths characteristic of those in the cirrus.
\item
CO was detected at more than 10 kpc away from the centers of NGC4438 and M86. For example, strong CO emission was found in 
the tidal structure north of NGC4438, in the regions E2E, E2, and E2W. Through these cloud complexes, 5$\times$10$^7$\msun\ or 9\%
of the gas in the center of NGC4438 could be lost to the ICM and potentially to neighboring galaxies. 
\item
CO was discovered in an \ha\ stream that ends in the center of M86, at $\sim$10 kpc southeast from the center, and near the peak of the 
\hi\ emission. The CO recession velocity agrees well with that of the \hi\ and the stars. The \cooz\ line properties indicate the 
presence of 2$\times$10$^7$ \msun\ of \htwo\ in the combined regions V3W and V3WW. Some of this gas could have formed in situ given 
that the local \hi\ surface brightness exceeds 10$^{20}$ cm$^{-2}$.
\item
The filamentary structure 11 kpc northeast from the center of M86, in the region that is denoted V7, contains CO moving at $-$530\kms . 
Its corresponding \htwo\ mass is 7$\times$10$^6$ \msun .  Because there is no associated \hi\ emission in this region or in its vicinity, 
we attribute this CO detection to gas that was primarily transferred to V7 in molecular form.  We interpret it as survival of  self-shielded, 
re-forming molecular clouds in filaments for $\sim$100 Myr. This is longer than the fragmentation timescale of individual isolated clouds, 
and shorter than the evaporation timescale due to heat conduction from the ICM.
\item
The integrated gas-to-dust ratio in the inner 2\arcmin\ of M86, 20-60, is too low to be consistent with gas cooling from the ICM. 
The molecular gas detected in and near M86 is therefore likely to have originated from an external galaxy.
\item
The most probable scenario for the molecular gas origin of M86 is that it comes from NGC4438. The best evidence for this remains the smooth 
\ha\ velocity gradient in the bridge between them, combined with the similar \ha\ and CO velocities, where detected.  
\item
The scenario of gas being transferred to M86 from NGC4388 on its south was also examined, given the \hi\ stream that is seen in projection 
between these galaxies. However, the non-detection of CO below or in the observed range of \hi\ recession velocities, i.e., between 
2000$-2700$\,\kms , does not favor this scenario.
\end{itemize}


\begin{acknowledgements}
 K. D. acknowledges support by the Centre National d\'\,Etudes Spatiales (CNES). We are thankful to G. Novak for constructive 
 discussions, to the HeVICS team for making their data publicly available, and to the anonymous referee who helped us to
 improve this document. Based on observations carried out with the  IRAM 30\,m telescope. IRAM is supported by INSU/CNRS 
 (France), MPG (Germany) and IGN (Spain).
\end{acknowledgements}


{}


\begin{thebibliography}{}

\bibitem[Bender et al.(1992)]{bender92}
Bender, R., Burstein, D., \& Faber, S. M. 1992, ApJ, 399, 462

\bibitem[Bloemen et al.(1986)]{bloemen86}
Bloemen, J. B. G. M., Strong, A. W., Mayer-Hasselwander, H. A., et al. 1986, A\&A, 154, 25	

\bibitem[Braine et al.(1997)]{braine97}
Braine, J., Henkel, C., \& Wiklind, T. 1997, A\&A, 321, 765 

\bibitem[Braine et al.(2000)]{braine00}
Braine, J., Lisenfeld, U., Duc, P.-A., Leon, S.  2000, Nature, 403, 867
	
\bibitem[Bregman \& Roberts(1990)]{bregman90}
Bregman, J. N., \& Roberts, M. S. 1990, ApJ 362, 468 

\bibitem[Cayatte et al.(1990)]{cayatte90}	
Cayatte, V., van Gorkom, J. H., Balkowski, C., \& Kotanyi, C., 1990 AJ, 100, 604

\bibitem[Chung et al.(2009)]{chung09}	
Chung, A., van Gorkom, J. H., Kenney, J. D. P., Crowl, H., \& Vollmer, B. 2009, AJ, 138, 1741

\bibitem[Clemens et al.(2010)]{clemens10}
Clemens, M. S., Jones, A. P., Bressan, A., et al. 2010, A\&A, 518, L50	
	
\bibitem[Cluver et al.(2010)]{cluver10}
Cluver, M. E., Appleton, P., Boulanger, F., et al. 2010, ApJ, 710, 248 

\bibitem[Colbert et al.(1996)]{colbert96}	
Colbert, E. J. M., Baum, S. A., Gallimore, J. F., O\'\,Dea, C. P., \& Christensen, J. A. 1996, ApJ, 467, 551

\bibitem[Combes et al.(1988)]{combes88}
Combes, F., Dupraz, C., Casoli, F., \& Pagani, L. 1988, A\&A, 203, L9 

\bibitem[Conselice et al.(2001)]{conselice01}
Conselice, C. J., Gallagher, J. S., III, \& Wyse, R. F. G. 2001, AJ, 122, 2281 

\bibitem[Cortese et al.(2010a)]{cortese10a}
Cortese, L., Bendo, G. J., Isaak, K. G., et al. 2010a, MNRAS, 403, L26

\bibitem[Cortese et al.(2010b)]{cortese10b}
Cortese, L., Bendo, G. J., Boselli, A., et al. 2010b, A\&A, 518, L63

\bibitem[Cowie \& McKee(1977)]{cowie77}
Cowie, L. L. \& McKee, C. F., 1977, ApJ, 211, 135

\bibitem[Cowie et al.(1983)]{cowie83}
Cowie, L. L., Hu, E. M., Jenkins, E. B., \& York, D. G. 1983, ApJ, 272, 29

\bibitem[Crowl\,et al.(2005)]{crowl05}
Crowl, H. H., Kenney, J. D. P., van Gorkom, J. H., \& Vollmer, B., 2005, AJ, 130, 65
	
\bibitem[Dalgarno \& McCray(1972)]{dalgarno72}
Dalgarno, A. \& McCray, R. A. 1972, ARA\&A, 10, 375

\bibitem[Davies et al.(2010)]{davies10}
Davies, J. I., Baes M., Bendo G.J., 2010, et al. A\&A 518, L48

\bibitem[Downes et al.(1993)]{downes93}
Downes, D., Solomon, P. M., \& Radford, S. J. E. 1993, ApJ, 414, L13	

\bibitem[Fabian(1994)]{fabian94}
Fabian, A. C. 1994, ARA\&A, 32, 277

\bibitem[Fabian et al.(2006)]{fabian06}
Fabian, A. C., Sanders, J. S., Taylor, G. B., et al. 2006, MNRAS, 366, 417
	
\bibitem[Ferland et al.(2008)]{ferland08}
Ferland, G. J., Fabian, A. C., Hatch, N. A., et al. 2008, 386, L72

\bibitem[Ferland et al.(2009)]{ferland09}
Ferland, G. J., Fabian, A. C., Hatch, N. A., et al. 2009, MNRAS, 392, 1475

\bibitem[Forman et al.(1979)]{forman79}
Forman, W., Schwarz, J., Jones, C., Liller, W., \& Fabian, A. C. 1979, ApJ, 234, L27

\bibitem[Gao \& Solomon(2004)]{gao04}
Gao, Y., \& Solomon, P. M. 2004, ApJ, 606, 271

\bibitem[Gavazzi et al.(2005)]{gavazzi05}
Gavazzi, G., Donati, A., Cucciati, O., et al. 2005, A\&A, 430, 411

\bibitem[Gomez et al.(2010)]{gomez10}
Gomez, H. L., Baes, M., Cortese, L., et al. 2010, A\&A, 518, L45

\bibitem[Gunn \& Gott(1972)]{gunn72}
Gunn, J. E., \& Gott, J. R. III 1972, ApJ, 176, 1

\bibitem[Heckman et al.(1989)]{heckman89}
Heckman, T. M., Baum, S. A., van Breugel, W. J. M., \& McCarthy, P. 1989, ApJ, 338, 48

\bibitem[Hollenbach \& Tielens(1997)]{hollenbach97}
Hollenbach, D. J., \& Tielens, A. G. G. M.  1997, ARA\&A, 35, 179
	
\bibitem[Hota et al.(2007)]{hota07}
Hota, A., Saikia, D. J., \& Irwin, J. A. 2007, MNRAS, 380, 1009

\bibitem[Irwin et al.(2000)]{irwin00}
Irwin, J. A., Saikia, D. J., \& English, J. 2000, AJ, 119, 1592
	
\bibitem[Johnstone et al.(2007)]{johnstone07}
Johnstone, R. M., Hatch, N. A., Ferland, G. J., et al. 2007, MNRAS, 382, 1246

\bibitem[Kenney \& Young(1989)]{kenney89}
Kenney, J. D. P., \& Young, J. S. 1989, ApJ, 344, 171
		
\bibitem[Kenney et al.(2008)]{kenney08}
Kenney, J. D. P.,  Tal T., Crowl H. H., et al. 2008, ApJ, 687, L69 

\bibitem[Kennicutt(1998)]{kennicutt98}
Kennicutt, R., C., Jr. 1998, ARA\&A, 36, 189

\bibitem[Krumholtz et al.(2011)]{krumholtz11}
Krumholz, M. R., Dekel, A., \& McKee, C. F.,  2011, ApJ, in press, arXiv:1109.4150
	
\bibitem[Li \& van Gorkom(2001)]{li01}
Li, Y., \& van Gorkom, J. 2001, ASP Conf 240, 637 

\bibitem[Liu \& Bregman(2005)]{liu05}
Liu, J.-F., \& Bregman, J. N., 2005, ApJS, 157, 59

\bibitem[O\'\,Sullivan et al.(2009)]{osullivan09}
O\'\,Sullivan, E. O., Giacintucci, S., Vrtilek, J. M., Raychaudhury, S., \& David, L. P. 2009, ApJ, 701, 1560

\bibitem[Oosterloo \& van Gorkom(2005)]{oosterloo05}
 Oosterloo, T., \& van Gorkom, J. 2005, A\&A, 437, L19

\bibitem[Machacek et al.(2004)]{machacek04}
Machacek, M. E., Jones, C., \& Forman, W. R.  2004, ApJ, 610, 183

\bibitem[Mahdavi et al.(1996)]{mahdavi96}
Mahdavi, A., Margaret J., Fabricant, D. G., et al. 1996, AJ, 111, 1

\bibitem[McKee \& Cowie(1977)]{mckee77}
McKee, C. F., \& Cowie, L. L. 1977, ApJ, 215, 213

\bibitem[Mei et al.(2007)]{mei07}
Mei, S., Blakeslee, J. P., Cot\'e, P., et al. 2007, ApJ, 655, 144

\bibitem[Mihos et al.(2005)]{mihos05}
Mihos, J. C., Harding, P., Feldmeier, J., \& Morrison, H.  2005, ApJ, 631, L41

\bibitem[Nulsen(1982)]{nulsen82}
Nulsen, P. E. J. 1982, MNRAS, 198, 1007

\bibitem[Papadopoulos \& Seaquist(1998)]{papadopoulos98}
Papadopoulos, P. P., \& Seaquist, E. R. 1998, ApJ, 492, 521

\bibitem[Randall et al.(2008)]{randall08}
Randall, S., Nulsen, P., Forman, W. R., et al. 2008, ApJ, 688, 208

\bibitem[Rangarajan et al.(1995)]{rangarajan95}	
Rangarajan, F. V. N., White, D. A., Ebeling, H., \& Fabian, A. C. 1995, MNRAS, 277, 1047	
	
\bibitem[Salom\'e et al.(2006)]{salome06}
Salom\'e, P., Combes, F., Edge, A. C., et al. 2006, A\&A, 454, 437 

\bibitem[Salom\'e et al.(2011)]{salome11}
Salom\'e, P.. Combes, F., Revaz, Y. et al. 2011, A\&A, 531, 85
	 
\bibitem[Schaye(2004)]{schaye04}
Schaye, J. 2004, ApJ, 609, 667

\bibitem[Solomon et al.(1997)]{solomon97}
Solomon, P. M., Downes, D., Radford, S. J. E., \& Barrett, J. W.  1997, ApJ, 478, 144

\bibitem[Sivanandam et al.(2010)]{sivanandam10}
Sivanandam, S., Rieke, M. J., \& Rieke, G. H. 2010, ApJ, 717, 147	
	
\bibitem[Smith et al.(2000)]{smith00}
Smith, R. J., Lucey, J. R., Hudson, M. J., Schlegel, D. J., \& Davies, R. L., 2000, MNRAS, 313, 469

\bibitem[Sun et al.(2010)]{sun10}
Sun, M., Donahue, M., Roediger, E., et al. 2010, ApJ, 708, 946	

\bibitem[Tamura et al.(2009)]{tamura09}
Tamura, T.; Maeda, Y.; Mitsuda, K., et al. 2009, ApJ, 705, L62	
	
\bibitem[Trinchieri et al.(2005)]{trinchieri05}
Trinchieri, G., Sulentic, J., Pietsch, W., \& Breitschwerdt, D. 2005, A\&A, 444, 697

\bibitem[Veilleux et al.(1999)]{veilleux99}
Veilleux, S., Bland-Hawthorn, J., Cecil, G., Tully, R. B., \& Miller, S. T. 1999, ApJ, 520, 11

\bibitem[Vollmer et al.(2001)]{vollmer01}
Vollmer, B., Cayatte, V., Balkowski, C., \& Duschl, W. J. 2001, ApJ, 561, 708	

\bibitem[Vollmer et al.(2005)]{vollmer05}
Vollmer, B., Braine, J., Combes, F., \& Sofue, Y. 2005, A\&A, 441, 473

\bibitem[Vollmer et al.(2008)]{vollmer08}
Vollmer, B., Braine, J., Pappalardo  C., \& Hily-Blan, P. 2008, A\&A 491, 455 

\bibitem[Vollmer et al.(2009)]{vollmer09}
Vollmer, B., Soida, M., Chung, A., et al. 2009 A\&A, 496, 669	

\bibitem[Willick et al.(1997)]{willick97}
Willick, J. A., Courteau, S., Faber, S. M., et al. 1997, ApJS, 109, 333

\bibitem[Wiklind et al.(1995)]{wiklind95}
Wiklind, T., Combes, F., \& Henkel, C. 1995, A\&A, 297, 643

\bibitem[Yoshida et al.(2004)]{yoshida04}
Yoshida, M., Ohyama, Y., Iye, M., et al. 2004, AJ, 127, 90

\bibitem[Young \& Scoville(1991)]{young91}
Young, J. S., \& Scoville, N. Z. 1991,ARA\&A, 29, 581
	
\end{thebibliography}
\end{document}